\documentclass[iop,apj,a4paper]{emulateapj}%published ApJ version
\usepackage{amsmath,amssymb,epsf}
\begin{document}

\title{Genuine Irregular Galaxies as a Relic of Building Blocks of Galaxies}

\author{K. Terao\altaffilmark{1},
Y. Taniguchi\altaffilmark{2},
M. Kajisawa\altaffilmark{1,2},
Y. Shioya\altaffilmark{2},
M. A. R. Kobayashi\altaffilmark{2},
K. Matsuoka\altaffilmark{4},
H. Ikeda\altaffilmark{1,5,6},
K. L. Murata\altaffilmark{3},
A. Ichikawa\altaffilmark{1},
M. Shimizu\altaffilmark{1},
M. Niida\altaffilmark{1}, and
E. Hamaguchi\altaffilmark{1}}

\altaffiltext{1}{Graduate School of Science and Engineering, Ehime
  University, Bunkyo-cho 2-5, Matsuyama, Ehime 790-8577, Japan}
\altaffiltext{2}{Research Center for Space and Cosmic Evolution, Ehime
  University, Bunkyo-cho 2-5, Matsuyama, Ehime 790-8577, Japan}
\altaffiltext{3}{Department of Particle and Astrophysical Science,
  Nagoya University, Furo-cho, Chikusa-ku, Nagoya 464-8602, Japan}
\altaffiltext{4}{Department of Physics and Astronomy, Seoul National
  University, 599 Gwanak-ro, Gwanak-gu, Seoul 151-742, Korea}
\altaffiltext{5}{Department of Astronomy, Kyoto University,
  Kitashirakawa-Oiwake-cho, Sakyo-ku, Kyoto 606-8502, Japan}
\altaffiltext{6}{Research Fellow of the Japan Society for the
  Promotion of Science}
\email{terao@cosmos.phys.sci.ehime-u.ac.jp}

\begin{abstract}

 In order to understand nature of building blocks of galaxies in the
 early universe, we investigate ``genuine irregular galaxies (GIGs)''
 in the nearby universe.  Here, GIGs are defined as isolated galaxies
 without regular structures (spheroid, bulge, disk, bar, spiral arm,
 and nucleus).  Using the results of two excellent studies on galaxy
 morphology based on the Sloan Digital Sky Survey (SDSS), we obtain a
 sample of 66 irregular galaxies.  We carry out new classification of
 them into GIGs and non-GIGs which have regular structure or show
 evidence for galaxy interaction, by using the SDSS Data Release 10
 images.  We then find that a half of these irregular galaxies (33/66)
 are GIGs and obtain an unambiguous sample of 33 GIGs for the first
 time.  We discuss their observational properties by comparing them
 with those of elliptical, S0, spiral galaxies, and irregular galaxies
 without the GIGs.  We find that our GIGs have smaller sizes, lower
 optical luminosities, bluer rest-frame optical colors, lower surface
 stellar mass densities, and lower gas metallicity than normal
 galaxies.  All these properties suggest that they are in chemically
 and dynamically younger phases even in the nearby universe.

\end{abstract}


%%%%%%%%%%%%%%%%%%%%%%%%%%%%%%%%%%%%%%%%%%%%%%%%%%%%%%%%%%%%%%%%%%%%%%%%%%%
\section{INTRODUCTION}

In the context of the hierarchical clustering scenarios for galaxy
evolution, all galaxies in the present day are considered to acquire
their masses by successive mergers of small building blocks of
galaxies (e.g., White \& Frenk 1991). Indeed, most galaxies observed
in high-redshift universe are generally much smaller and much less
massive than the present galaxies (Ferguson et al. 2004; van Dokkum et
al. 2008; Taniguchi et al. 2009).  Therefore, in order to understand
the formation and early evolution of galaxies, it is necessary to
explore which types of building blocks of galaxies were present and
how the mass assembly process proceeded in the early universe.

Here a question arises as ``are there any relics of such building
blocks in the nearby universe?''.  The best candidates are both gas-rich
dwarf galaxies and irregular galaxies because these galaxies have a
large amount of gaseous components even now\footnote{See the
proceedings of IAU Symposium Vol.255, ``Low Metallicity Star
Formation: From the First Stars to Dwarf Galaxies'' ed. L. K. Hunt,
S. Madden, \& R. Schneider (Cambridge: Cambridge Univ. Press)} (e.g.,
Gallagher \& Hunter 1984; Hunter \& Gallagher 1986; Hunter et
al. 2012).  Therefore, in this paper, we focus on irregular galaxies
as probable relic of building blocks of galaxies.

Since the galaxy morphology tells us dynamical properties of galaxies,
morphological classifications of galaxies provide us an important
guideline in the understanding of galaxies.  The most famous and
pioneering work was made by Hubble (1926, 1936): the so-called Hubble
classification of galaxy morphology.  In his work, he stated that a
few percent of galaxies do not show rotational symmetry and thus they
cannot be classified in his classification scheme.  They are called as
irregular galaxies (Irr).  He also stated that a half of irregular
galaxies including Magellanic Clouds constitute an important
population but the remaining galaxies are peculiar ones mostly due to
galaxy interaction.  Accordingly, irregular galaxies were recognized
as heterogeneous populations (e.g., Sandage 1961).

Since irregular galaxies are minor populations, most studies have been
devoted to the understanding of normal elliptical and spiral galaxies
to date.  Yet, irregular galaxies are important populations if they
are not related to any galaxy interaction and mergers.  Here we call
them ``genuine irregular galaxies (GIGs)'' in this paper.  GIGs are
defined as isolated irregular galaxies without any regular structures
(spheroid, bulge, disk, bar, spiral arm, and nucleus).  In particular,
we remind that the absence of nucleus is an important property,
already noted by Hubble (1936).  It is known that almost all ordinary
galaxies have a galactic nucleus, that is believed to harbor a
supermassive black hole (SMBH; Kormendy \& Richstone 1995; Kormendy \&
Ho 2013 and references therein).  The mass of SMBH is observed to be
well correlated to that of the spheroidal component (bulge or
spheroid) of galaxies (e.g. Magorrian et al. 1998; Kormendy \& Ho
2013).  Therefore the absence of nucleus implies that the spheroidal
component has not yet developed in that galaxy.

It is thus suggested that irregular galaxies are still in an early
evolutionary phase.  This is also advocated by the presence of a
plentiful amount of gas (e.g., Roberts \& Hayes 1994; Hunter \&
Elmegreen 2004).  Moreover, since irregular galaxies are less massive
down to $\sim 10^9~M_\odot$, we regard them as relics of building
blocks of galaxies.

Motivated by this idea, we define GIGs, select a sample of GIGs, and
then investigate their fundamental properties for the first time in
this paper.  Throughout this paper, magnitudes are given in the AB
system.  We adopt a flat universe with $\Omega_\mathrm{M} = 0.3$,
$\Omega_\Lambda = 0.7$, and $H_0 = 70~\mathrm{km~s^{-1}~Mpc^{-1}}$,
throughout the paper.

%%%%%%%%%%%%%%%%%%%%%%%%%%%%%%%%%%%%%%%%%%%%%%%%%%%%%%%%%%%%%%%%%%%%%%%%%%%
\section{Genuine Irregular Galaxies}

\subsection{Definition}

Originally, the class of irregular galaxies was defined by Hubble
(1926), who investigated morphological properties of $\sim 400$
galaxies in the nearby universe.  He stated that; {\it About 3 per
cent of the extra-galactic nebulae lack both dominating nuclei and
rotational symmetry.  These form a distinct class which can be termed
``irregular''.  The Magellanic Clouds are the most conspicuous
examples}.  This definition is our guideline and, indeed, is the same
as that of the GIGs discussed in this paper.  The reason why we use
the new term, GIG, is to exclude any contamination from the so-called
peculiar galaxies mostly made by galaxy interactions and mergers
(Hubble 1936).  Such contamination also includes active galaxies with
a superwind activity\footnote{These galaxies were once classified as
Irr II (e.g., Sandage 1961).} (e.g., M82) and jets from an active
galactic nucleus (e.g., M87).

To avoid any contamination, we re-define the class of GIGs as isolated
galaxies without any regular structure.  Here, the term of
``isolated'' means that there is no neighboring galaxy and there is no
firm evidence for a recent galaxy merger.  Note that the majority of
ultraluminous infrared galaxies such as Arp 220 look isolated galaxies
but they are major-merger remnants (e.g., Sanders \& Mirabel 1996).

\subsection{Samples}

In order to obtain a well-defined sample of GIGs, we need reliable
imaging survey for a large number of galaxies.  We also need
spectroscopic information to investigate their activity, metallicity,
and physical sizes.  One of the most suitable surveys for galaxies is
the Sloan Digital Sky Survey (SDSS; York et al. 2000).

In fact, the following two excellent studies have been already made on
morphological properties of galaxies:
\begin{enumerate}
 \item Fukugita et al. (2007; hereafter F07)\\ Based on the SDSS Data
       Release Three (DR3), they investigated 2253 galaxies with
       extinction-corrected $r$ magnitude brighter than 16~mag in the
       north equatorial stripe (230~deg$^2$).  A total of 1866
       galaxies also have spectroscopic information.  In their
       analysis, irregular galaxies are termed as Im (i.e., Magellanic
       irregular).  Among the 2253 galaxies, they identified 31 Im
       galaxies.  This provides the fraction of irregular galaxies is
       $\approx 1.4$ percent ($= 31 / 2253$).
 \item Nair \& Abraham (2010; hereafter NA10)\\ Based on the SDSS Data
       Release Four (DR4), they investigated 14034 galaxies with
       extinction-corrected $g$ magnitude brighter than 16~mag in all
       SDSS fields (6670~deg$^2$).  All of their sample galaxies have
       spectroscopic information and lie in the redshift range, $0.01
       < z < 0.1$.  Among the 14034 galaxies, they identified 35 Im
       galaxies.  This provides the fraction of irregular galaxies is
       $\approx 0.25$ percent ($= 35 / 14034$).  Note that NA10 also
       obtained 52 Sdm and 68 Sm galaxies in their analysis.  However,
       since these galaxies show evidence for spiral arms, we use only
       35 Im galaxies in the later discussion.
\end{enumerate}
Then, we obtain 66 irregular galaxies (31 from F07 and 35 from NA10)
in total.  We list up these galaxies in Tables~\ref{tb:ListOfIrrsF07}
(F07) and \ref{tb:ListOfIrrsNA10} (NA10), separately.  These 66
galaxies are our preliminary sample of irregular galaxies.

It is noted that there is no overlap of Im galaxies between F07 and
NA10 and the number fraction of the Im galaxies to the total galaxies
is different with each other.  This seems due to the following
reasons.  First, the data source is different between F07 (DR3) and
NA10 (DR4).  Second, the survey area of F07 is restricted to the north
equatorial stripe (230~deg$^2$), while that of NA10 is all SDSS fields
(6670~deg$^2$), which includes the survey area of F07.  Third, the
selection band is different: $r \leq 16$ (F07) and $g \leq 16$ (NA10).
Fourth, NA10 uses a spectroscopic catalog and adopts the redshift
criterion, $0.01 < z < 0.1$, while F07 uses a photometric catalog
which includes the galaxies without spectroscopic information.

Much attention has been paid to dwarf irregular galaxies and very
faint irregular galaxies because they can be more probable fossils of
building blocks of galaxies (e.g., Hunter et al. 2012; Brown et
al. 2013).  However, we do not take absolute magnitudes into account
in our sample selection because our main aim is to identify GIGs
regardless of their luminosity.

\subsection{Selection of GIGs}

In order to identify unambiguous GIGs from the preliminary sample of
66 irregular galaxies, we examine their optical images carefully by
using the SDSS Data Release Ten (DR10) database.  Here we use the
eye-inspection method. All the authors have examined individual
galaxies and then have re-classified the preliminary sample into the
following classes:
\begin{enumerate}
 \item[(1)] Elliptical-like galaxies:\\ No object is found.
 \item[(2)] Disk-like galaxies:
 \begin{enumerate}
  \item[(2-a)] The presence of bulge with both spiral arms and a
	       bar:\\  No object is found.
  \item[(2-b)] The presence of bulge with spiral arms and without
	       bar:\\  Four objects are found.  Their color
	       montages constructed by assigning RGB colors $g$, $r$,
	       $i$ data channels\footnote{The images are taken from
	       the SDSS Imaging Server:
	       http://skyserver.sdss3.org/dr10/en/tools/chart/navi.aspx}
	       are shown in Figure~\ref{fig:Disk+Bulge+Arm-Bar}.
	       \begin{figure*}
		\epsscale{.96}
		\plotone{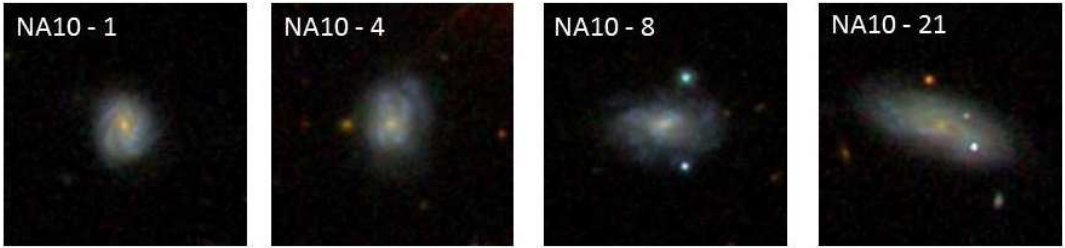}
		\caption{
		Montage of images of the 4 non-GIGs which are
		re-classified as disk with bulge and spiral arms and
		without a bar taken from the SDSS DR10.  North is up
		and east is left.  Image sizes are
		$60~\mathrm{arcsec} \times 60~\mathrm{arcsec}$.
		}
		\label{fig:Disk+Bulge+Arm-Bar}
	       \end{figure*}
  \item[(2-c)] The presence of bulge without spiral arms and with a
	       bar:\\  No object is found.
  \item[(2-d)] The absence of bulge with both spiral arms and a bar:\\
	       One object is found.  Its image is shown in
	       Figure~\ref{fig:Disk-Bulge+Arm+Bar}.
	       \begin{figure*}
		\epsscale{0.35}
		\plotone{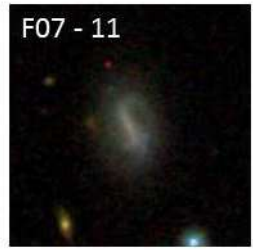}
		\caption{
		Same as Figure~\ref{fig:Disk+Bulge+Arm-Bar} but for
		the 1 non-GIG which are re-classified as disk without
		bulge and with both spiral arm and a bar.
		}
		\label{fig:Disk-Bulge+Arm+Bar}
	       \end{figure*}
  \item[(2-e)] The presence of bulge without both spiral arms and a
	       bar:\\  Six objects are found.  Their images are
	       shown in Figure~\ref{fig:Disk+Bulge-Arm-Bar}.
	       \begin{figure*}
		\epsscale{.96}
		\plotone{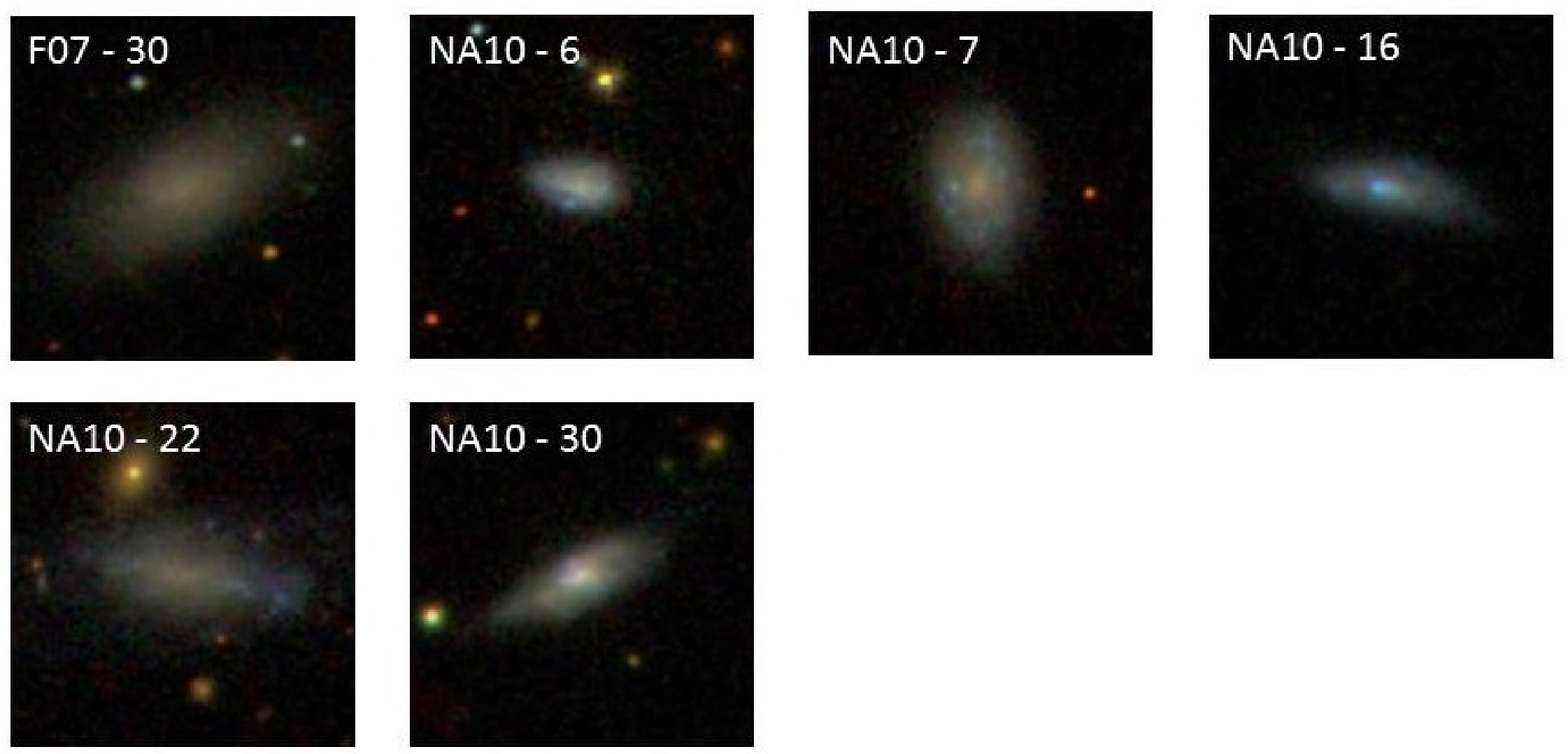}
		\caption{
		Same as Figure~\ref{fig:Disk+Bulge+Arm-Bar} but for
		the 6 non-GIGs which are re-classified as disk with
		bulge and without both spiral arm and a bar.
		}
		\label{fig:Disk+Bulge-Arm-Bar}
	       \end{figure*}
  \item[(2-f)] The absence of bulge with spiral arms and without a
	       bar:\\  No object is found.
  \item[(2-g)] The absence of bulge without spiral arms and with a
	       bar:\\  No object is found.
 \end{enumerate}
 \item[(3)] The presence of a partner:\\ Eight objects are found.
	    Their images are shown in Figure~\ref{fig:Interacting}.
	    All these galaxies appear to be interacting galaxies.
	    Comments on the individual galaxies are given in
	    Appendix~\ref{sec:appA}.
	    \begin{figure*}
	     \epsscale{1.0}
	     \plotone{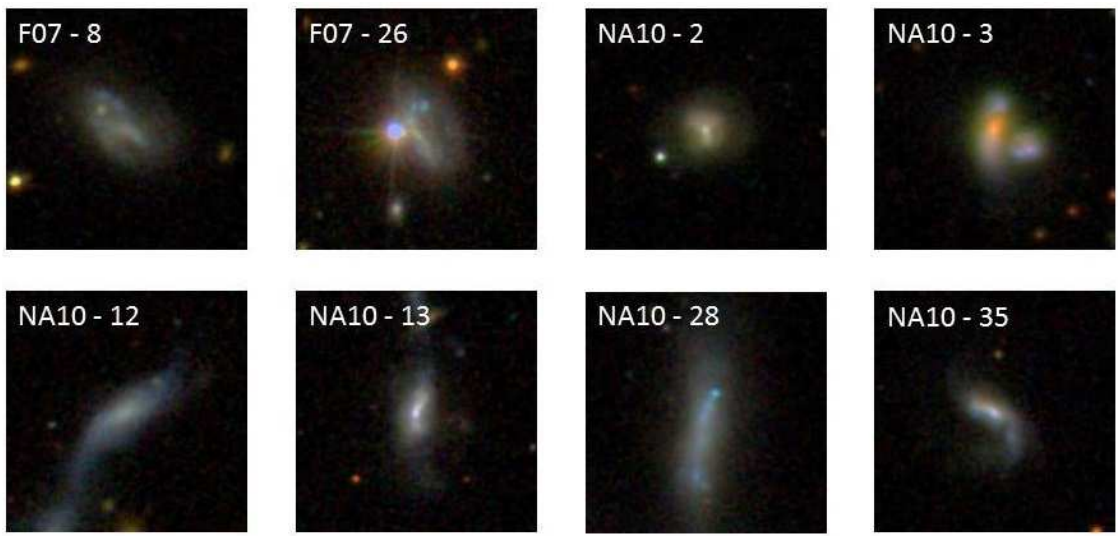}
	     \caption{
	     Same as Figure~\ref{fig:Disk+Bulge+Arm-Bar} but for the 8
	     non-GIGs which are re-classified as interacting galaxies,
	     non-GIG/I.
	     }
	     \label{fig:Interacting}
	    \end{figure*}
 \item[(4)] Merger remnants:\\  Fourteen objects are found.  Their
	    images are shown in Figure~\ref{fig:Merging}.  All these
	    galaxies appear to be merging galaxies.  Comments on the
	    individual galaxies are also given in
	    Appendix~\ref{sec:appA}.
	    \begin{figure*}
%	     \epsscale{1.3} % for Shioya-san's environment (Windows)
	     \epsscale{1.02} % for Terao's or MARK's enrironment (Linux)
	     \plotone{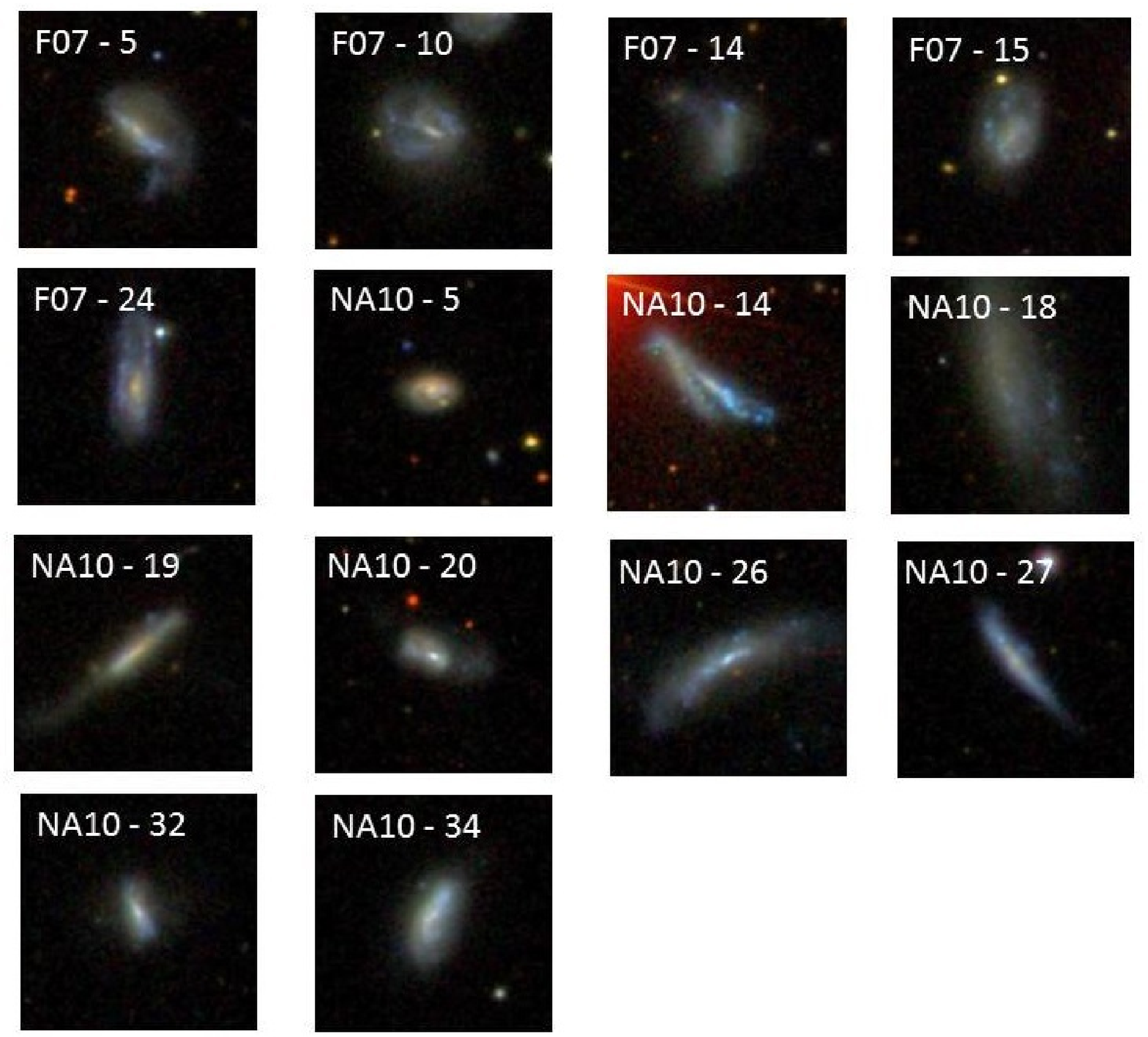}
	     \caption{
	     Same as Figure~\ref{fig:Disk+Bulge+Arm-Bar} but for the
	     14 non-GIGs which are re-classified as merging galaxies,
	     non-GIG/M.
	     }
	     \label{fig:Merging}
	    \end{figure*}
 \item[(5)] GIGs:\\  Thirty three objects are found.  Their images
	    are shown in Figure~\ref{fig:GIGs}.
	    \begin{figure*}
	     \epsscale{1.1}
	     \plotone{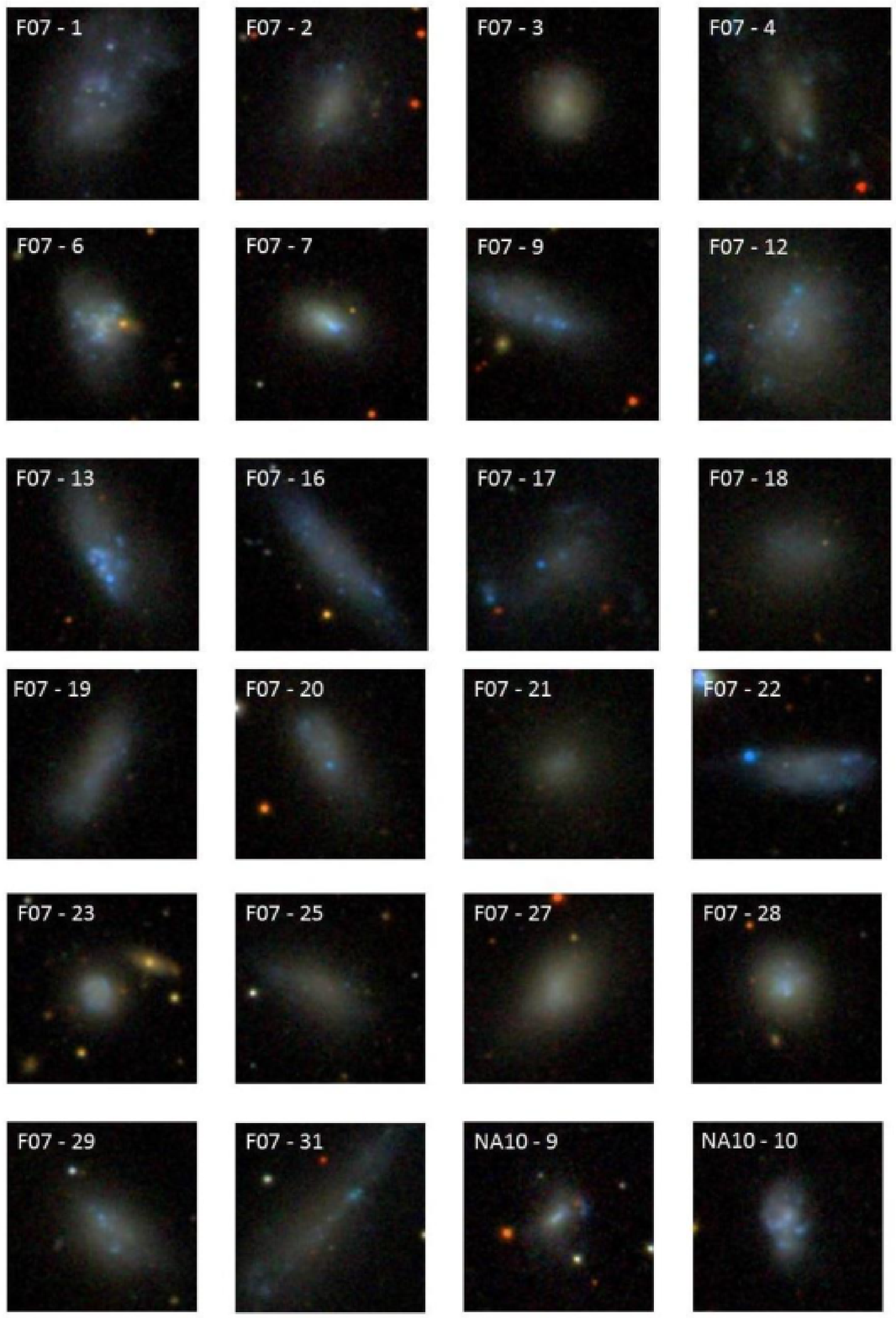}
	     \caption{
	     Same as Figure~\ref{fig:Disk+Bulge+Arm-Bar} but for the
	     33 GIGs.
	     }
	     \label{fig:GIGs}
	    \end{figure*}
	    \setcounter{figure}{5}
	    \begin{figure*}
	     \epsscale{1.08}
	     \plotone{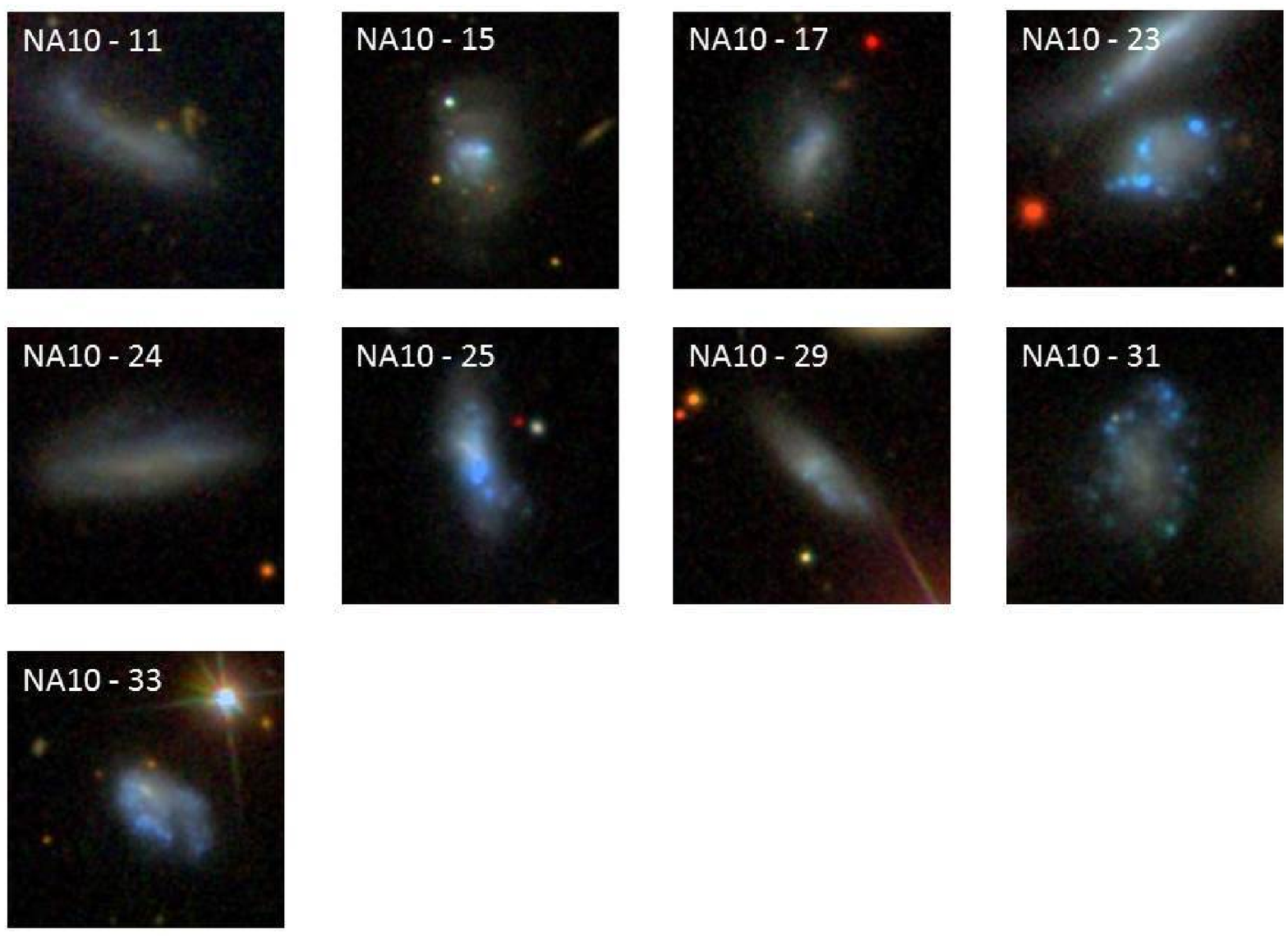}
	     \caption{
	     (Continued.)
	     }
	    \end{figure*}
	    Although four objects (F07-22, F07-23,
	    NA10-23, and NA10-29) seem to have companion in their
	    images, they can be regarded as isolated as described
	    in Appendix~\ref{sec:appB}.
\end{enumerate}

We find that only 33 galaxies out of the 66 preliminary samples of
irregular galaxies are re-classified into the unambiguous GIGs (22
from F07 and 11 from NA10).  The remaining 33 galaxies are not GIGs
(hereafter, non-GIGs).  The sub-classes of the non-GIGs re-classified
into (1), (2), (3), and (4) above are dubbed as non-GIG/E, non-GIG/Sp,
non-GIG/I, and non-GIG/M, respectively.  In
Table~\ref{tb:ListOfNonGIGs}, we summarize the statistics of the
non-GIGs.

The redshifts are available for 28 GIGs (17 from F07 and 11 from NA10)
among the 33 GIGs. The redshift information on F07-GIGs is taken from
the SDSS DR10 database.  However, it is noted that the redshift of one
of the GIGs, F07-25, is given as 0.8124, which is extremely higher
than the redshifts of the other F07-GIGs with spectroscopic redshifts
($z \sim 0.0032$--0.027)\footnote{If this redshift for F07-25 were
  correct, its absolute $g$-band magnitude $M_g$ and size
  $R_\mathrm{P}$ would be estimated as $M_g = -27.28$ and
  $R_\mathrm{P} = 156.6$~kpc, respectively.}.  Since its spectrum
appears to be very noisy, we do not include this object in later
discussion.  Therefore our final spectroscopic sample consists of 27
GIGs (16 from F07 and 11 from NA10); hereafter we call them as zGIGs.
We summarize the observational and physical information of the 33 GIGs
including 27 zGIGs in Table~\ref{tb:ListOfGIG}.

We find that only a half of the preliminary samples of irregular
galaxies are identified as unambiguous GIGs.  Among the 33 non-GIGs,
33 percent ($= 11 / 33$) is re-classified into the disk-like galaxies
(non-GIG/Sp) while the remaining 67 percent ($= 22 / 33$) is either
interacting or merging galaxies (non-GIG/I$+$M).  It is unclear
whether or not there are any intrinsic differences between the
non-GIGs and the GIGs.  In order to answer the question, here we show
comparisons of the $g$-band absolute magnitudes $M_g$ and surface
brightnesses $\mu_g$ of the non-GIGs with redshift information (31 out
of 33 non-GIGs) and the zGIGs.

\subsection{Comparison between GIGs and Non-GIG Samples}

\begin{figure*}
 \epsscale{1.}
 \plottwo{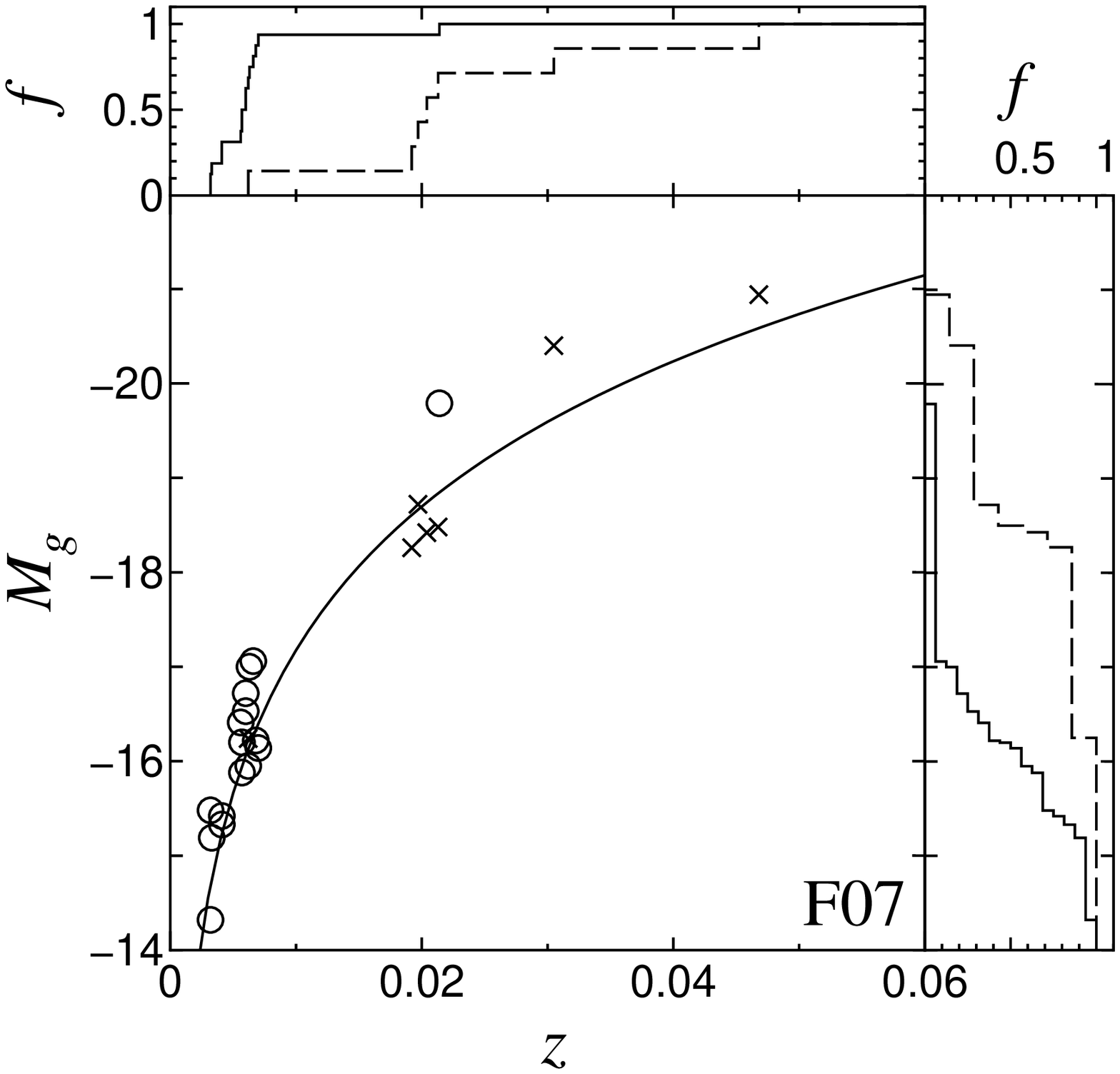}{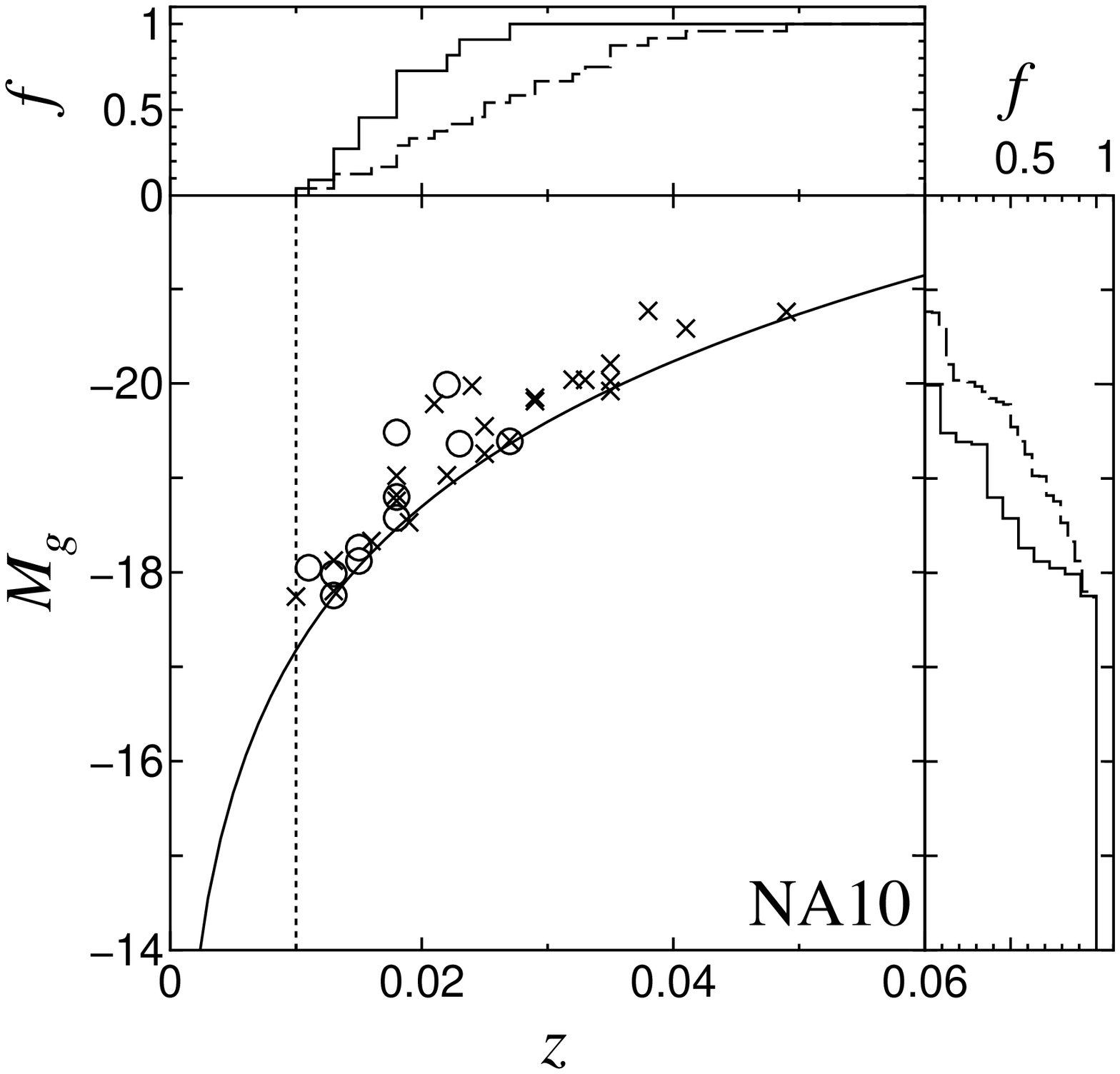}
 \caption{
 A comparison of the $g$-band absolute magnitude $M_g$ and redshift
 distributions for F07 (left) and NA10 (right).
 In the main panels, the zGIGs and non-GIGs are
 represented by the open circles and the crosses, respectively.  The
 solid curves present a constant $g$-band magnitude, $m_g = 16$, which
 is the selection criterion in NA10.  The vertical dotted-line in the
 right main panel is the threshold redshift adopted in NA10, $z =
 0.01$.  The plot in the main panel is projected onto the two side
 panels where a cumulative histogram is displayed for each population
 in each of the dimensions.  In the two sub-panels, the zGIGs and
 non-GIGs are shown by the solid and dashed histograms, respectively.
 }
 \label{fig:Mg+z}
\end{figure*}

The zGIGs are fainter than the non-GIGs as clearly shown in the
right-hand side cumulative histograms in the main panels of
Figure~\ref{fig:Mg+z}.  We also find that there is no GIGs with $M_g
\lesssim -20$ in both of F07 and NA10 samples.  Among the very faint
galaxies with $M_g \gtrsim -17$, which are available only in the F07
samples\footnote{The selection criteria of $m_g < 16$ and $z > 0.01$
adopted in NA10 exclude these faint galaxies as shown in
Figure~\ref{fig:Mg+z}.}, all galaxies except for a non-GIG, F07-30,
are the zGIGs.  Are these differences intrinsic one or merely a
consequence of a kind of selection effect?

\begin{figure*}
 \epsscale{1.}
 \plottwo{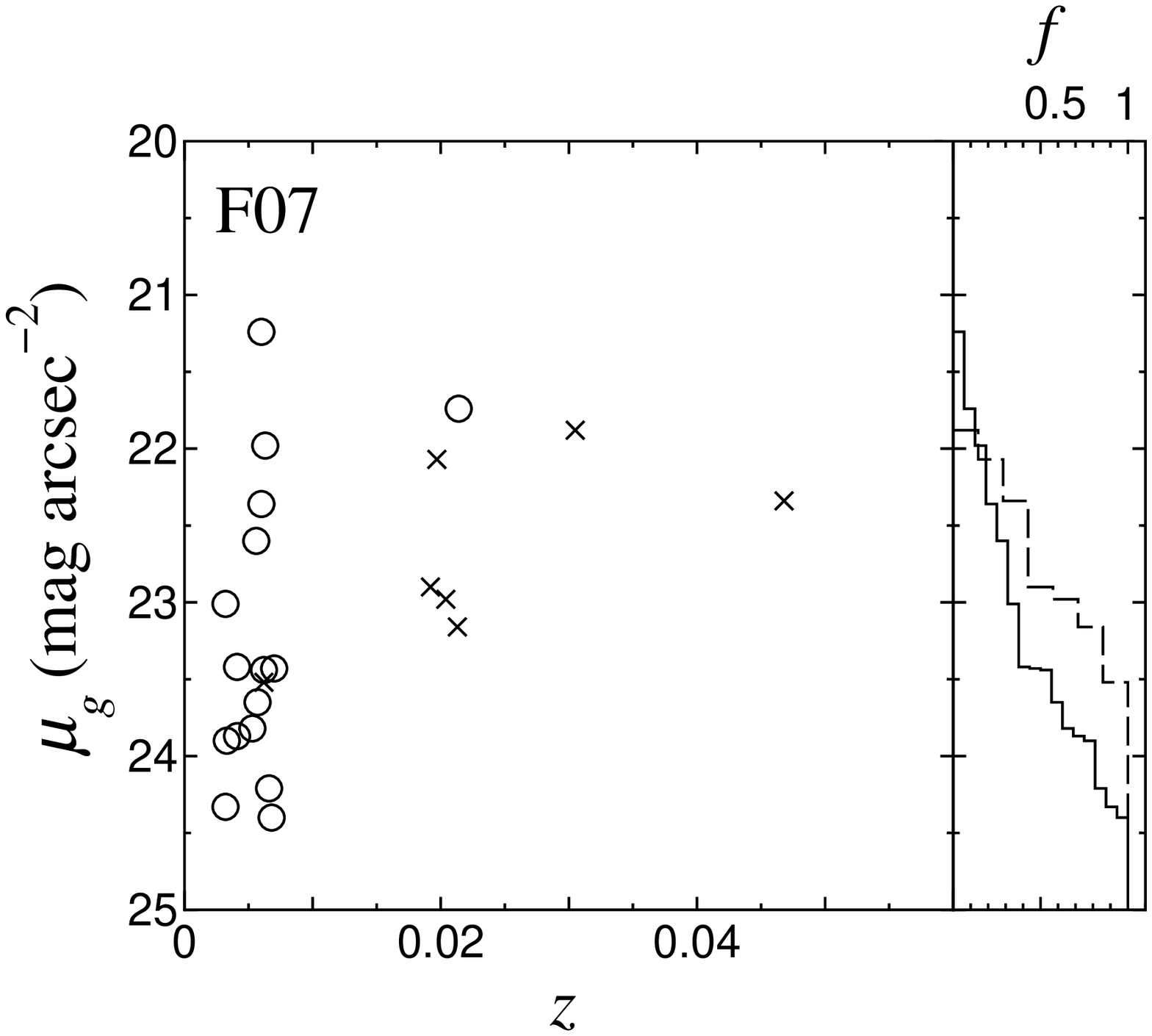}{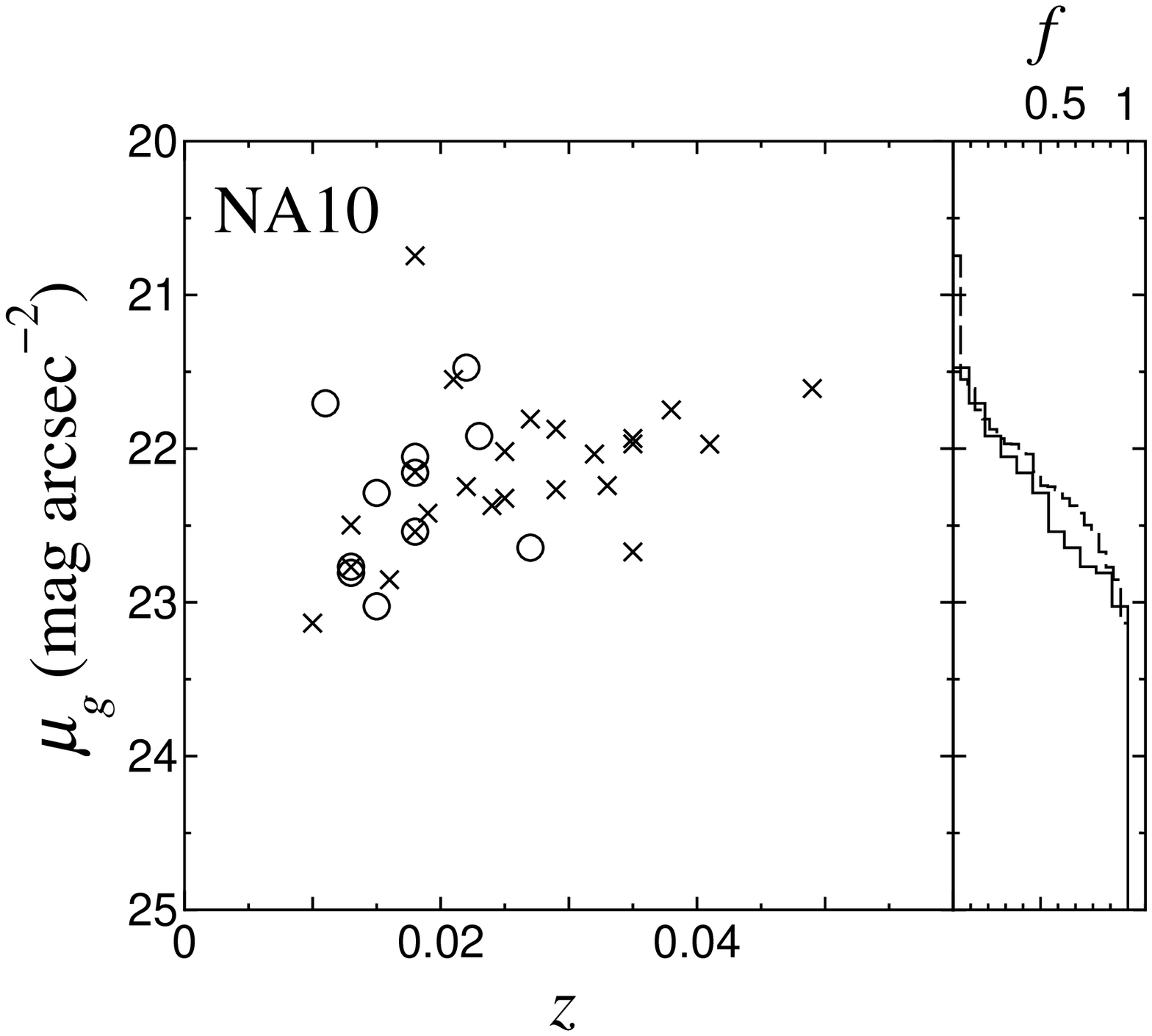}
 \caption{
 Same as Figure~\ref{fig:Mg+z} but for $g$-band surface brightness
 $\mu_g$ and redshift distributions for F07 (left) and NA10 (right).
 }
 \label{fig:mu_g+z}
\end{figure*}

As shown in Figure~\ref{fig:mu_g+z}, the cumulative histograms of
$\mu_g$ for the NA10-zGIGs and non-GIGs are similar with each other.
Almost half of the F07-zGIGs ($= 9 / 16$) have $\mu_g \lesssim
23.5~\mathrm{mag~arcsec^{-2}}$, which is similar to the non-GIGs.
Therefore, we can conclude that these zGIGs are bright enough to find
any signatures of the presence of spiral arm, bar, and/or
interaction/merger if exist.  On the contrary, the remaining half of
the F07-zGIGs ($= 7 / 16$) is fainter than
$23.5~\mathrm{mag~arcsec^{-2}}$, which is comparable to that of the
low surface brightness galaxies (LSBGs; e.g., Impey et al. 1996),
although the GIGs do not have disk by definition.  Hence, for these
faint-$\mu_g$ zGIGs (from the faintest in $\mu_g$, F07-17, F07-21,
F07-31, F07-9, F07-22, F07-19, and F07-20), it should be reminded
there remains a possibility that they have signatures of spiral arm,
bar, and/or interaction/merger.

Including the GIGs without zGIGs, about 30 percent ($= 10 / 33$) are
so faint in $g$-band surface magnitude (i.e., $\mu_g >
23.5~\mathrm{mag~arcsec^{-2}}$) that we might miss any signatures of
regular structure and/or interaction/merger.  Nevertheless, the
remaining 70\% of the GIGs ($= 23 / 33$) are sufficiently bright in
$\mu_g$ and hence they are regarded as unambiguous GIGs.


%%%%%%%%%%%%%%%%%%%%%%%%%%%%%%%%%%%%%%%%%%%%%%%%%%%%%%%%%%%%%%%%%%
\section{Nature of the GIGs}

In the previous section, we have newly defined the morphological class
of GIGs as isolated galaxies without any regular structures.  Then we
selected a sample of 33 GIGs based on the two papers on the
morphological classification of galaxies by using the SDSS data.
Among the 33 GIGs, 27 have reliable redshift information, which are
dubbed as zGIGs.

In this section, we investigate observational properties of the 27
zGIGs through comparisons with normal galaxies.  As a control sample
of normal galaxies that includes elliptical (E), S0, and spiral (Sp)
galaxies, we use the SDSS galaxies studied by NA10 because all their
galaxies have spectroscopic information and the sample number is
significantly larger than that of F07.  The total numbers of E/S0 and
Sp galaxies are 5938 and 7708, respectively.

All the rest-frame photometric data for the NA10 galaxies are taken
from Table~1 in NA10.  The rest-frame photometry for F07 galaxies with
redshift information is estimated by using the SDSS DR10 data with the
$k$-correction following the procedure shown in Chilingarian \&
Zolotukhin (2012).  Note also that all the photometric data are
corrected for the Galactic extinction.

%%%%%%%%%%%%%%%%%%%%%%%%%%%%%%%%%%%%%%%%%%%%%%%%%%%%%%%%%%%%%%%%%%
\subsection{$g$-band Luminosity Function}\label{subsec:gLF}

\begin{figure*}
 \epsscale{1.}
 \plottwo{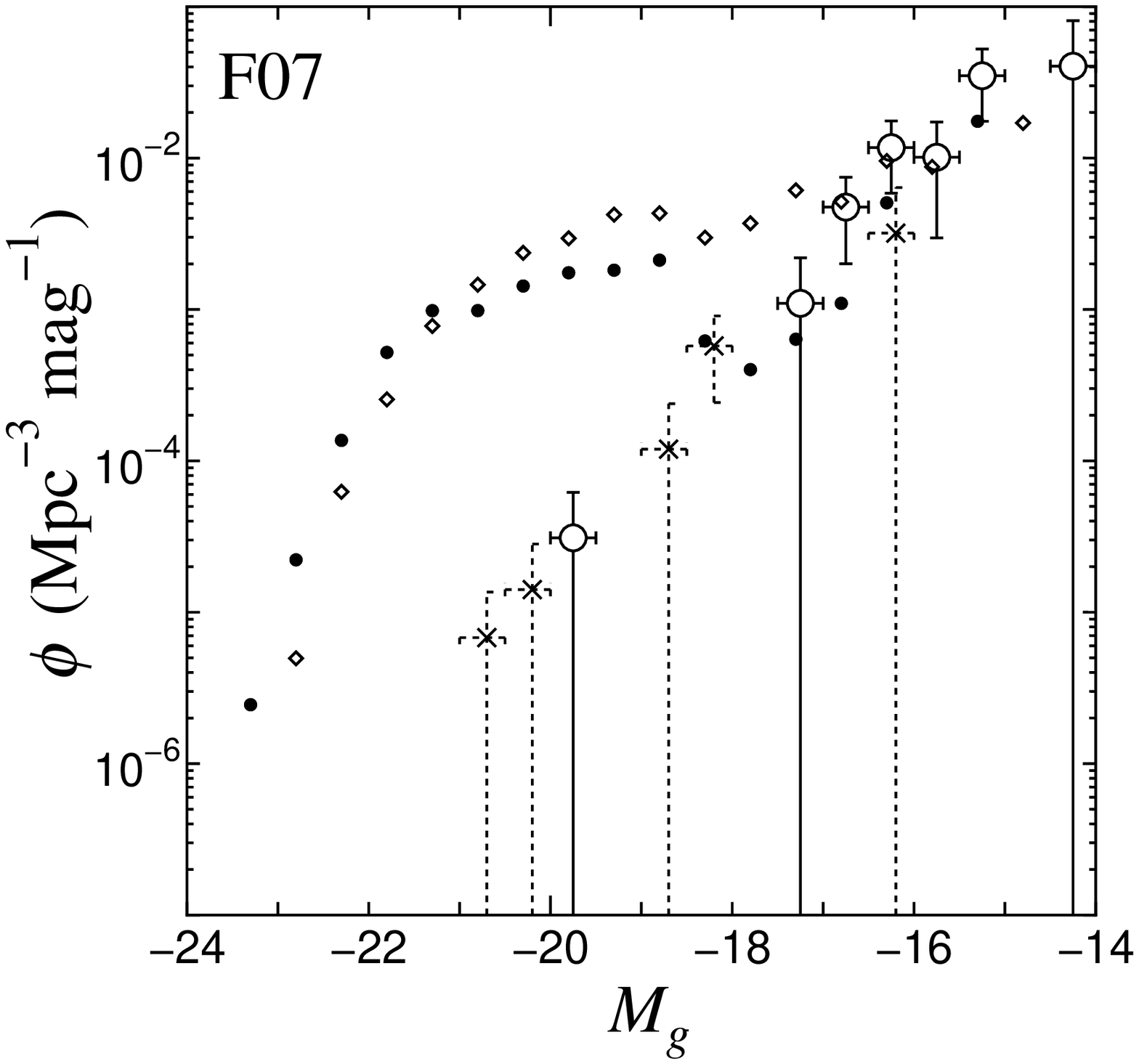}{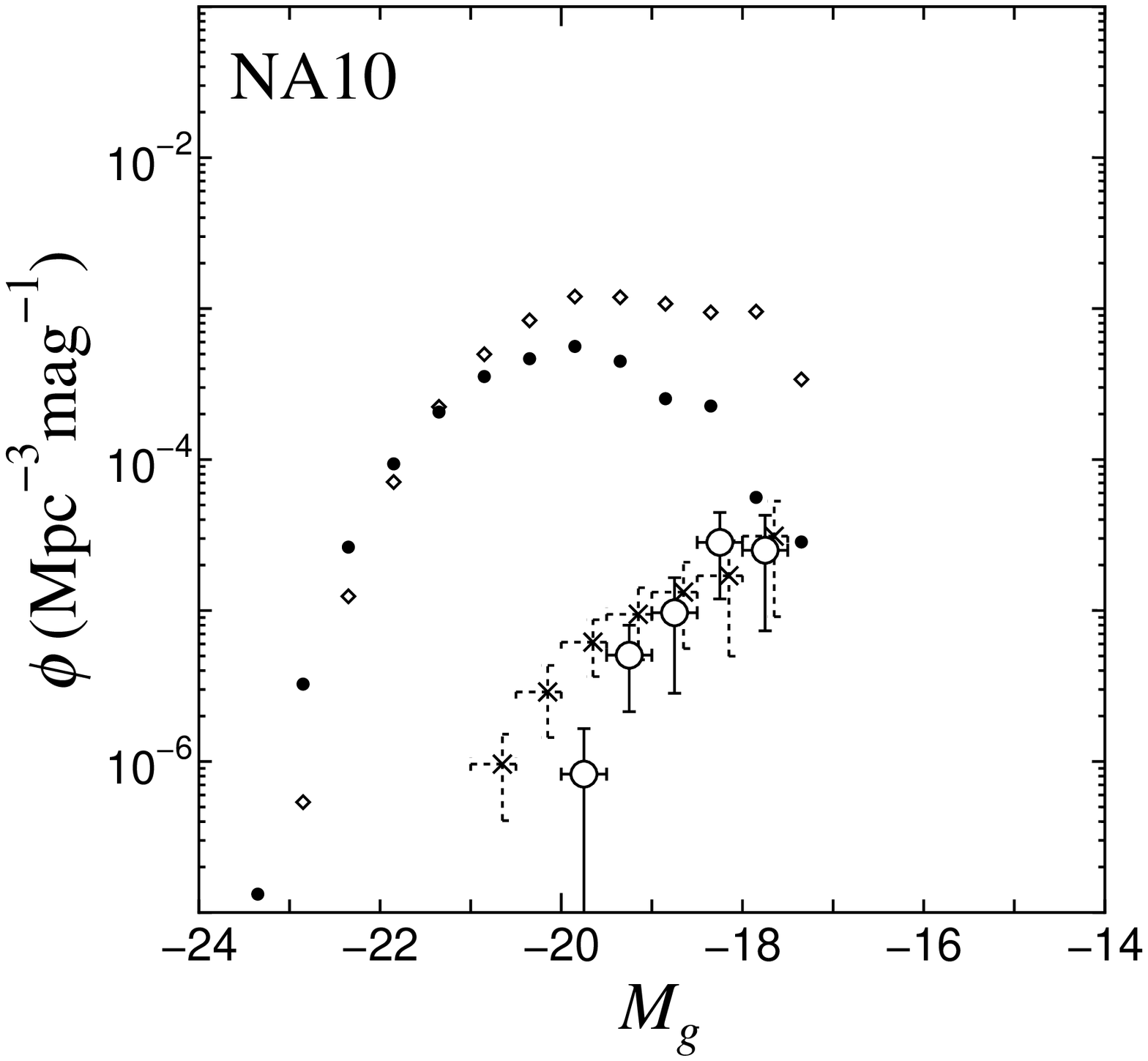}
 \caption{
 The $g$-band luminosity functions for F07 (left) and NA10 (right) GIGs.
 The $g$-band luminosity functions of GIGs, non-GIGs, E/S0, and spiral
 (Sp) galaxies are shown by open circles, crosses, small filled
 circles, and small open diamonds, respectively.  The 1$\sigma$
 Poissonian errors of the number densities are presented as vertical
 error-bars only for the GIGs and non-GIGs because of their small
 numbers.  To avoid an overlap of symbols, the symbols of E/S0
 galaxies, spiral galaxies, and non-GIGs are shifted horizontally by
 $-0.1$, $-0.1$, and $+0.1$~mag, respectively.
 }
 \label{fig:MgLF}
\end{figure*}

We compare the number densities of the zGIGs to the control sample of
normal galaxies and non-GIGs in order to examine whether or not the
small detection number of the zGIGs implies intrinsically small number
density.  We show the luminosity functions (LF) of GIGs together with
those of E/S0, Sp, and non-GIGs in Figure~\ref{fig:MgLF}.  The
LFs are estimated by the so-called $V_{\rm max}$ method (Schmidt 1968).
Note, however, that the correction for the completeness is not made.
The results are shown for the F07 and NA10 samples separately in
Figure~\ref{fig:MgLF}.  In this Figure, the results are shown for the
F07 and NA10 samples, separately.

First, we find that the LFs of GIGs show power-law like LFs for both
the F07 and NA10 samples.  Second, we find that the GIGs are the
dominant population in fainter galaxies at $M_g > -17$.  On the other
hand, the fraction of GIGs is negligible ($< 1\%$) at $M_g < -19$.

%%%%%%%%%%%%%%%%%%%%%%%%%%%%%%%%%%%%%%%%%%%%%%%%%%%%%%%%%%%%%%%%%%
\subsection{Sizes and Absolute $g$-band Magnitudes}

\begin{figure*}
 \epsscale{1.}
 \plotone{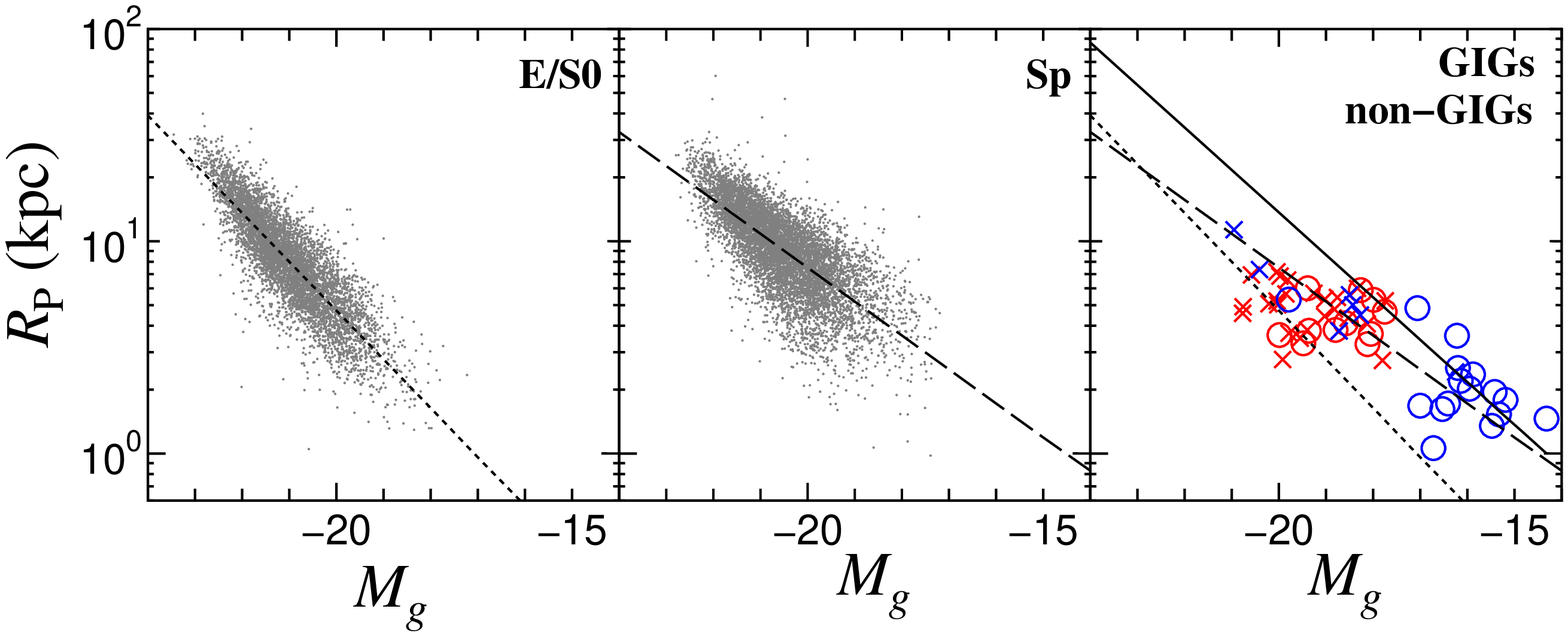}
 \caption{
 Diagrams between the size $R_\mathrm{P}$ and the $g$-band absolute
 magnitude $M_g$ for E/S0 (left), Sp (middle), and zGIGs and non-GIGs
 with redshift information (right).  The regression lines for E/S0 and
 Sp are shown by the dotted and dashed lines, respectively.  These
 lines are also shown in the right panel for reference.  In the right
 panel, blue and red symbols are for the F07 and NA10 samples,
 respectively.  A locus of constant $g$-band surface brightness of
 $\mu_g = 23.5~\mathrm{mag~arcsec^{-2}}$ is also shown by solid line
 for reference.
 }
 \label{fig:Rp-Mg}
\end{figure*}

We compare the physical size of zGIGs as a function of $M_g$ with
those of normal galaxies and the non-GIGs.  Here, we use the Petrosian
radius, $R_\mathrm{P}$, as a size of galaxy.  The results are shown in
Figure \ref{fig:Rp-Mg}.  Compared to the control samples, the sizes of
the zGIGs are significantly smaller.  However, the $R_{\rm P}$-$M_g$
relation for the GIGs appears to follow the same relation for the Sp
sample.

The sizes of NA10-zGIGs (F07-zGIGs without F07-25 at $z = 0.8$) are
3.28--6.01~kpc (1.06--2.20~kpc) with a median value of 3.82~kpc
(1.68~kpc).  We find that the F07-zGIGs are systematically smaller
than the NA10-zGIGs.  This can be attributed to the fact that the
F07-zGIGs are located at lower redshifts than the redshift threshold
adopted in NA10 ($z = 0.01$).

%%%%%%%%%%%%%%%%%%%%%%%%%%%%%%%%%%%%%%%%%%%%%%%%%%%%%%%%%%%%%%%%%%
\subsection{Optical Colors}

\begin{figure*}
 \epsscale{1.}
 \plotone{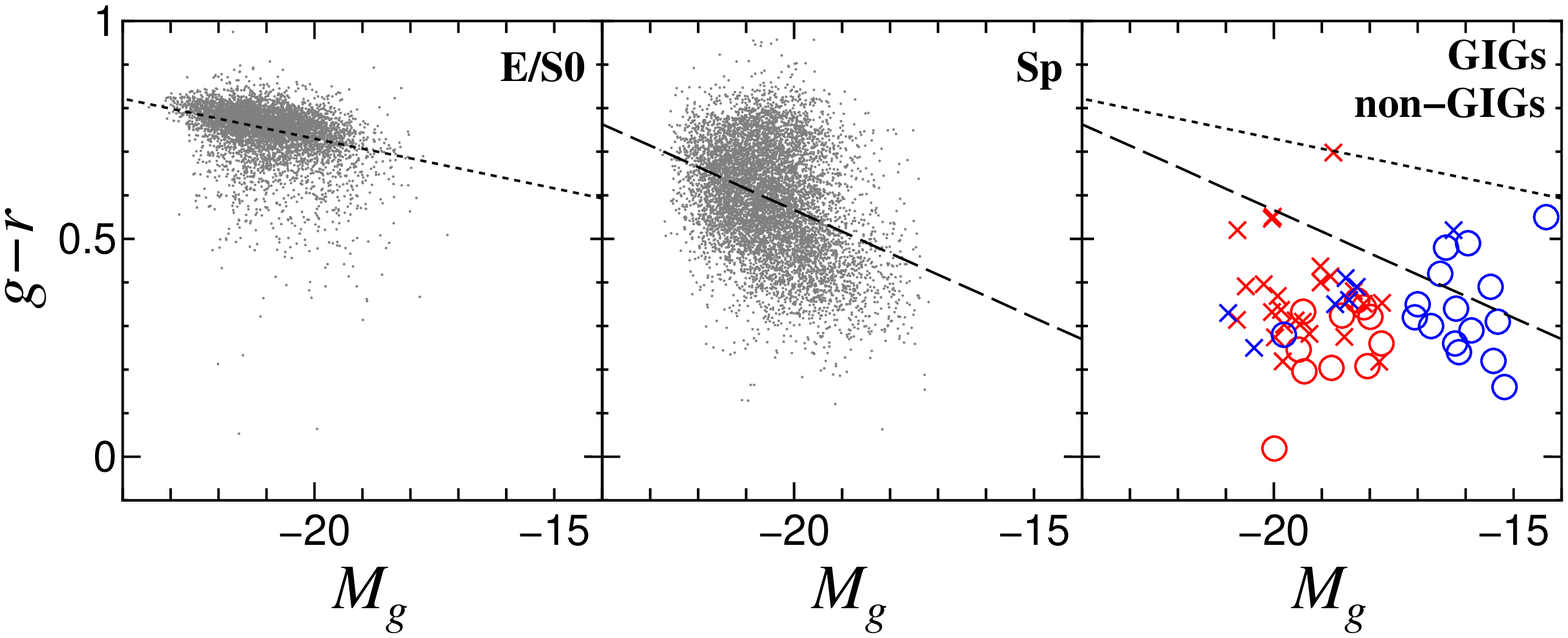}
 \caption{
 Same as Figure~\ref{fig:Rp-Mg} but for the rest-frame $g-r$ color and
 the $g$-band absolute magnitude.
 }
 \label{fig:CMR}
\end{figure*}

We compare the rest-frame optical color, $(g-r)_0$, of zGIGs with
those of normal galaxies as a function of $M_g$ in
Figure~\ref{fig:CMR}.  The zGIGs tend to be bluer than normal spiral
galaxies on average and all the zGIGs are bluer than $(g - r)_0 =
0.6$.  If the bluer color of zGIGs is attributed to recent massive
star formation events, their star formation rate relative to the
stellar mass (i.e., the specific star formation rate) is
systematically higher than those of normal spiral galaxies (see
Section~\ref{subsec:IonSource}).

%%%%%%%%%%%%%%%%%%%%%%%%%%%%%%%%%%%%%%%%%%%%%%%%%%%%%%%%%%%%%%%%%%
\subsection{Stellar Mass Density}

\begin{figure*}
 \epsscale{1.}
 \plotone{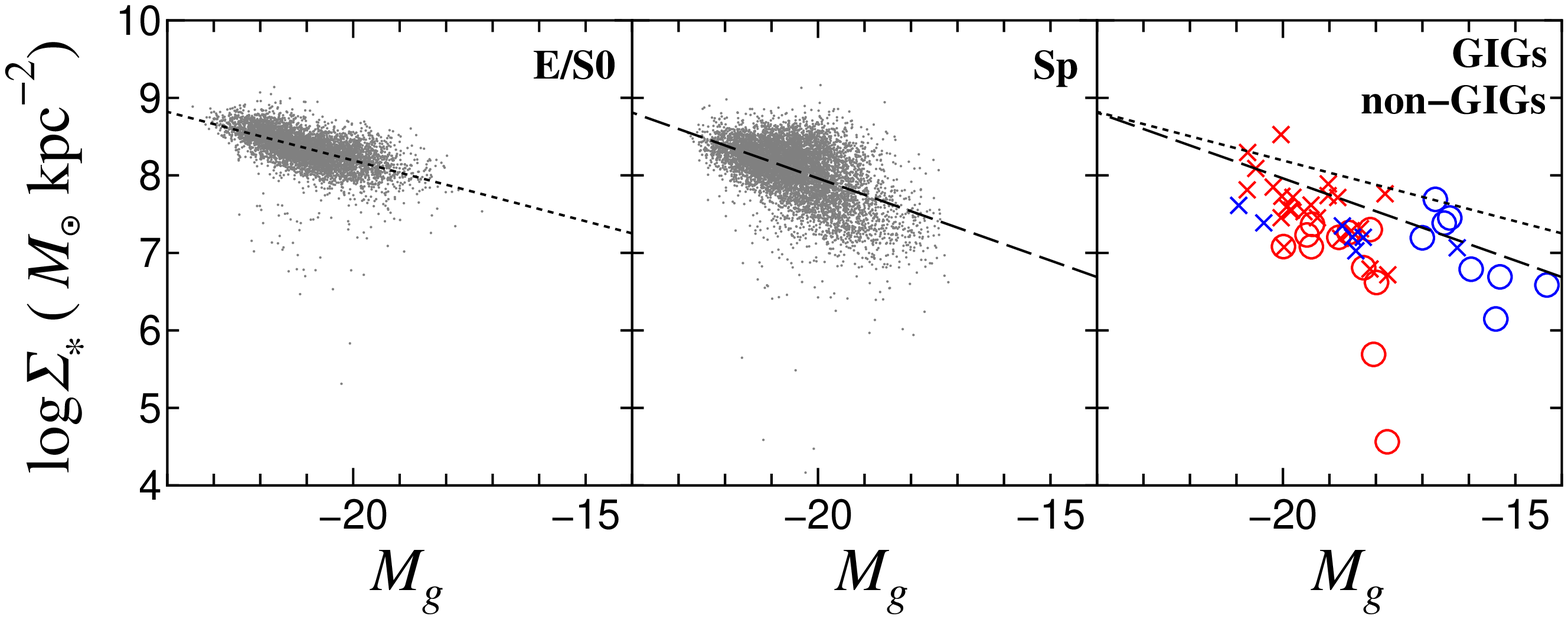}
 \caption{
 Same as Figure~\ref{fig:Rp-Mg} but for the surface mass density
 $\Sigma_\ast$ and the $g$-band absolute magnitude
 }
 \label{fig:muM-Mg}
\end{figure*}

We compare the stellar mass density, $\Sigma_*$, of the zGIGs with
normal galaxies as a function of $M_g$ in Figure~\ref{fig:muM-Mg}.
The catalog of NA10 includes stellar mass estimated by 
Kauffmann et al. (2003a).  For zGIGs in F07, we use the stellar mass
in the same catalog.  The GIGs tend to have lower stellar mass density
compared to the Sp sample systematically.  This suggests that the
cumulative star formation rate in the GIGs is smaller than that of
typical spiral galaxies.

%%%%%%%%%%%%%%%%%%%%%%%%%%%%%%%%%%%%%%%%%%%%%%%%%%%%%%%%%%%%%%%%%%
\subsection{Ionization Sources}\label{subsec:IonSource}

\begin{figure*}
 \epsscale{.5}
 \plotone{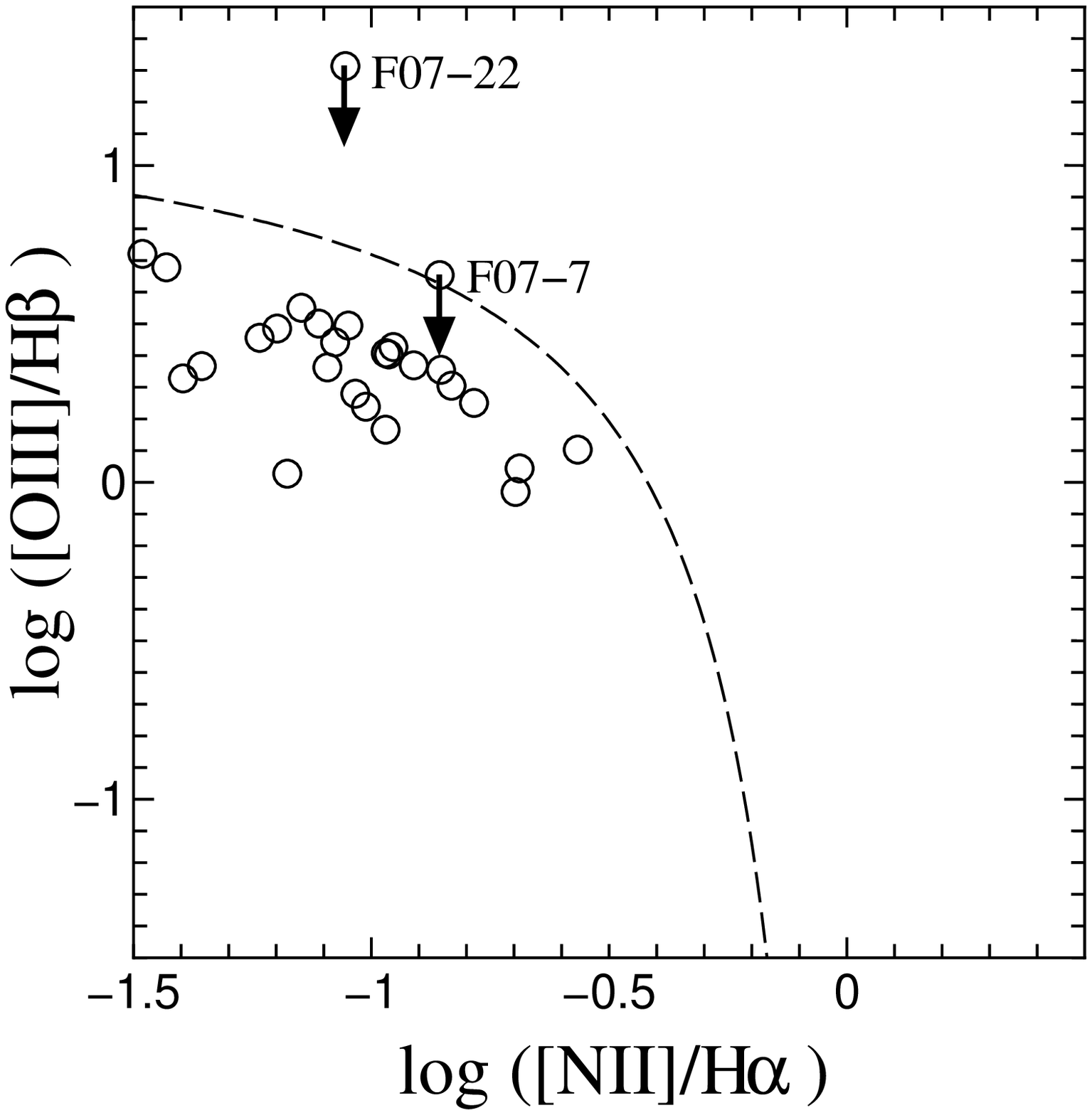}
 \caption{
 Diagram between $[\mbox{\ion{N}{2}}]\lambda 6584 / \mathrm{H}\alpha$
 and $[\mbox{\ion{O}{3}}]\lambda 5007 / \mathrm{H}\beta$ for the 27
 zGIGs, for which optical spectra are available in the SDSS data
 archive.  The dashed curve shows the border between
 $\mbox{\ion{H}{2}}$ region and AGN excitation taken from Kauffmann et
 al. (2003b).  For the two objects, F07-7 (SDSS~J$105248.63+000203.9$)
 and F07-22 (SDSS~J$131743.19-010003.8$), the upper limits of the
 $[\mbox{\ion{O}{3}}]\lambda 5007 / \mathrm{H}\beta$ ratio are shown
 because strong H$\beta$ absorption appears in their spectra.  It is
 thus considered that the apparent very high
 $[\mbox{\ion{O}{3}}]\lambda 5007 / \mathrm{H}\beta$ ratio is due to
 the underestimate of H$\beta$ emission.
 }
 \label{fig:BPT}
\end{figure*}

All the 27 zGIGs show strong optical emission lines.  In order to
study major ionization sources in them, we use the so called
excitation diagram between $[\mbox{\ion{N}{2}}]\lambda 6584 /
\mathrm{H}\alpha$ and $[\mbox{\ion{O}{3}}]\lambda 5007 /
\mathrm{H}\beta$ emission line ratios (e.g., Baldwin, Phillips, \&
Terlevich 1981; Veilleux \& Osterbrock 1987).  Since the GIGs have no
nucleus by definition, there is no contribution from active galactic
nuclei.  It is therefore expected that most probable ionization
sources are massive OB stars.  In fact, we find that all the zGIGs
show $\mbox{\ion{H}{2}}$-region like properties in
Figure~\ref{fig:BPT}.  Here, the emission line flux data are taken
from the SDSS DR10 datebase.  Note that the two GIGs, F07-7
(SDSS~J$105248.63+000203.9$) and F07-22 (SDSS~J$131743.19-010003.8$),
show strong H$\beta$ absorption.  Since this makes difficult to
estimate their H$\beta$ emission, we show these data points as upper
limits in Figure~\ref{fig:BPT}.

We then conclude that the major ionization sources in the zGIGs are
massive OB stars.  If the superwind activity works effectively,
ionization would be dominated by shock heating.  If this is the case,
the zGIGs would be located around $[\mbox{\ion{N}{2}}]\lambda 6584 /
\mathrm{H}\alpha \sim 0$ (e.g., Veilleux \& Osterbrock 1987).
Therefore, it is suggested that the superwind activity is fairly low
in the zGIGs.

Finally, it should be mentioned that the majority of zGIGs are located
in a domain with both $[\mbox{\ion{N}{2}}]\lambda 6584 /
\mathrm{H}\alpha < -1$ and $[\mbox{\ion{O}{3}}]\lambda 5007 /
\mathrm{H}\beta > 0$ in Figure~\ref{fig:BPT}.  This implies that the
zGIGs are metal poor galaxies (e.g., Kewley et al. 2001). In fact,
typical nuclear starburst galaxies are located in a domain with both
$[\mbox{\ion{N}{2}}]\lambda 6584 / \mathrm{H}\alpha \sim -0.5$ and
$[\mbox{\ion{O}{3}}]\lambda 5007 / \mathrm{H}\beta < 0.5$ (Balzano
1983; Ho et al. 1997; Kauffmann et al. 2003b).  Based on the
theoretical study by Kewley et al. (2001), it is suggested that the
gas metallicity of zGIGs range from 0.5 $Z_{\odot}$ to 1 $Z_{\odot}$
(see next section).

%%%%%%%%%%%%%%%%%%%%%%%%%%%%%%%%%%%%%%%%%%%%%%%%%%%%%%%%%%%%%%%%%%
\subsection{Gas Metallicity}

Optical emission-line ratios are also used to estimate the gas
metallicity of galaxies (e.g., Kewley \& Ellison 2008 and references
therein).  In the estimate of gas metallicity, the so-called $R_{23}$
parameter has been often used: $R_{23} = ([\mbox{\ion{O}{2}}]\lambda
3727 + [\mbox{\ion{O}{3}}] \lambda 4959, 5007) / \mathrm{H}\beta$.
However, the $[\mbox{\ion{O}{2}}] \lambda 3727$ is outside of the SDSS
spectral coverage for most nearby galaxies.  Since it has been shown
that the $N2$ parameter, $N2 = \log{([\mbox{\ion{N}{2}}] \lambda 6584
/ \mathrm{H}\alpha)}$, also works well for the estimate of gas
metallicity (e.g., Denicol{\'o} et al. 2002; Pettini \& Pagel 2004),
we use the $N2$ parameter in our analysis.  We use the following $N2$
calibration formula obtained by Pettini \& Pagel (2004):

\begin{equation}
12 + \log (\mathrm{O/H}) = 8.90 + 0.57 \times N2.
\end{equation}

The gas metallicity obtained from this formula is given in the last
column of Table 3. The gas metallicity ranges from 8.08 to 8.81 with
the median value of 8.31.  Here we note that the gas metallicity of
F07-21, 8.81, appears to be unusually high for its absolute magnitude,
$M_g = -14.32$. Since its optical spectrum shows very strong H$\beta$
absorption, the H$\alpha$ emission flux could be underestimated
significantly. In order to estimate its gas metallicity unambiguously,
we need new detailed optical spectroscopy for this object.  It is also
encouraged to use other metallicity indicators, such as $R_{23}$, for
future investigations because there is a possible systematic
difference of the estimated gas metallicity from $R_{23}$ and $N2$
parameters, $\sim 0.3$~dex (e.g., Cullen et al. 2013).

\begin{figure*}
 \epsscale{.5}
 \plotone{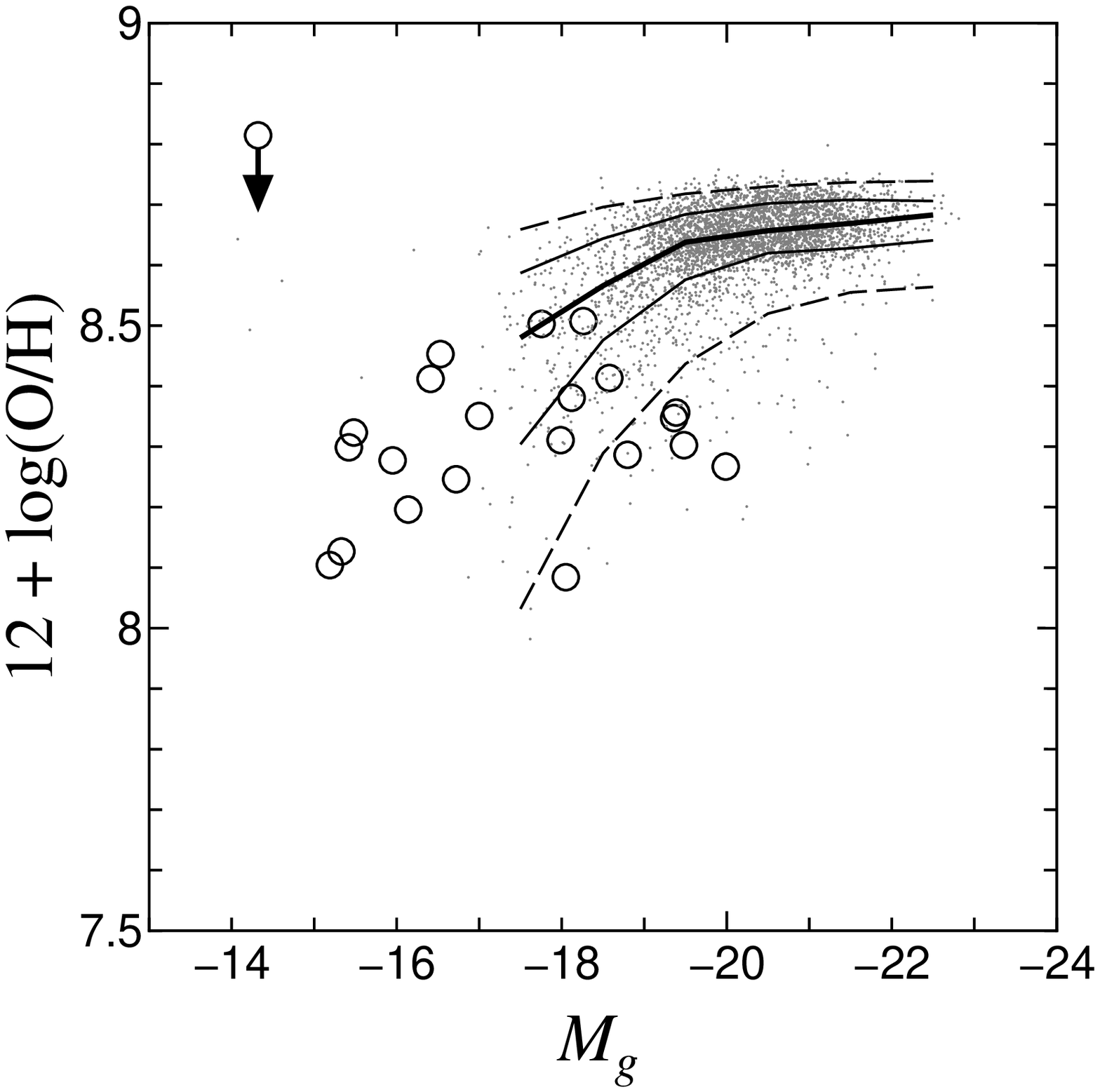}
 \caption{
 Comparison of gas metallicity derived from the N2 index given in
 equation (1) as a function of $g$-band absolute magnitude between the
 zGIGs and star-forming galaxies in NA10.  The zGIGs are represented
 by the open circles, while the SDSS star-forming galaxies are shown
 by dots.  The solid and dashed curves enclose 68\% and 95\% of them,
 respectively.
 }
 \label{fig:MMR}
\end{figure*}

In Figure~\ref{fig:MMR}, we show the gas metallicity to the absolute
$g$-band magnitude relation between the zGIG and the star-forming
galaxies in the NA10 catalog.  The metallicities of both samples are
evaluated from $N2$ parameter.  For the star-forming galaxies in NA10,
we use galaxies in which both H$\alpha$ and $[\mbox{\ion{N}{2}}]
\lambda 6584$ are detected in high significance ($S/N > 3$).  We also
show the contours that enclose 68\% and 95\% of the star-forming
galaxies as solid lines and dashed lines.  All zGIGs but F07-21 appear
to be systematically metal poorer than the star-forming galaxies in
NA10.  It is also noted that the gas metallicity of our zGIGs is lower
than the solar metallicity, $12 + \log{\mathrm{(O/H)}} = 8.66$
(Asplund et al. 2005), if we exclude F07-21.

%%%%%%%%%%%%%%%%%%%%%%%%%%%%%%%%%%%%%%%%%%%%%%%%%%%%%%%%%%%%%%%
\section{CONCLUSION}

In this paper, we have defined the class of genuine irregular galaxies
(GIGs) and selected a sample of 33 GIGs from literature.  This sample
is not a complete one in any sense.  Our main purpose is to
unambiguously pick up irregular galaxies originally defined by Hubble
(1926).  As noted by Hubble (1936), a half of irregular galaxies are
contaminations from interacting/merging galaxies and starburst
galaxies and galaxies with an AGN.  In fact, our analysis shows that
33 irregular galaxies among the 66 ones are not genuine irregular
galaxies.  It is thus suggested that $\sim 50$ percent of irregular
galaxies in previous statistical studies on galaxy morphology are not
GIGs and thus statistical properties of irregular galaxies are
affected by this selection bias.

More importantly, if we have a reliable sample of GIGs in the nearby
universe, we are able to examine their observational properties and
then obtain some observational clues in the understanding of building
blocks of galaxies.  In fact, our GIGs have smaller physical sizes and
absolute luminosities than those of normal galaxies.  Also, they have
no evidence for the bulge component.  This suggests that an SMBH is
not present even now since we cannot find their galactic nucleus.
Furthermore, since the GIGs have lower metallicity than that of normal
galaxies, they are in younger stages in the context of chemical
evolution of galaxies.

In conclusion, it is important to make systematic studies of GIGs,
providing us what happened in the early evolution of galaxies at high
redshift universe.  Our future plans are the followings.  (1) We will
make our own systematic surveys of GIGs using archival imaging data
obtained with the Suprime-Cam on the 8.2~m Subaru Telescope.  Our main
target fields are the SDF (Kashikawa et al. 2004), SXDS (Furusawa et
al. 2008) and COSMOS (Scoville et al. 2007; Taniguchi et al. 2007).
(2) In this paper, we have realized that eye-ball classification of
galaxy morphology is basically a hard business.  This means that we
need an automated and more sophisticated classification scheme of
GIGs.  We will prepare this type of software and then apply it to
future studies of galaxy morphology.

%%%%%%%%%%%%%%%%%%%%%%%%%%%%%%%%%%%%%%%%%%%%%%%%%%%%%%%%%%%%%%%%%%%%%%%

\acknowledgments

We would like to thank Takashi Murayama for useful discussion.  This
work was financially supported in part by the Japan Society for the
Promotion of Science (JSPS; No.23244031 [YT] and No. 23740152 [MK]).
HI is financially supported by the JSPS through the JSPS Research
Fellowship.
%%%%%%%%%%%%%%%%%%%%%%%%%%%%%%%%%%%%%%%%%%%%%%%%%%%%%%%%%%%%%%%%%%%%%%%

\setcounter{table}{0}
\begin{table*}
 \begin{center}
  \caption{A list of 31 irregular galaxies in F07.}
  \label{tb:ListOfIrrsF07}
  \begin{tabular}{lrccccl}
   \hline \hline
   \multicolumn{2}{c}{Name} & & \multicolumn{2}{c}{Observed Magnitude} &
   \colhead{$\mu_g$} & \colhead{Our Class\tablenotemark{a}}\\
   \cline{1-2} \cline{4-5}
   \colhead{Ours} & \colhead{F07} & & \colhead{$g$} & \colhead{$r$}
   & $(\mathrm{mag~arcsec^{-2}})$ & \\
   \hline
   F07-1  &   22 & & 15.47 & 15.23 & 23.00 & GIG \\
   F07-2  &   26 & & 16.14 & 15.78 & 23.42 & (z)GIG \\
   F07-3  &   41 & & 15.96 & 15.44 & 22.36 & (z)GIG \\
   F07-4  &   45 & & 16.79 & 16.14 & 23.44 & (z)GIG \\
   F07-5  &   47 & & 15.80 & 15.41 & 21.88 & non-GIG/M\\
   F07-6  &  402 & & 15.25 & 14.91 & 21.74 & (z)GIG \\
   F07-7  &  596 & & 15.51 & 15.17 & 21.24 & (z)GIG \\
   F07-8  &  658 & & 16.55 & 16.09 & 23.16 & non-GIG/I\\
   F07-9  &  709 & & 15.70 & 15.50 & 23.90 & (z)GIG \\
   F07-10 &  744 & & 15.98 & 15.61 & 22.11 & non-GIG/M\\
   F07-11 &  773 & & 16.50 & 16.08 & 22.90 & non-GIG/Sp\\
   F07-12 &  861 & & 15.32 & 14.91 & 23.01 & (z)GIG \\
   F07-13 &  952 & & 15.93 & 15.82 & 23.27 & GIG \\
   F07-14 & 1030 & & 16.41 & 16.04 & 22.98 & non-GIG/M\\
   F07-15 & 1046 & & 16.04 & 15.68 & 22.07 & non-GIG/M\\
   F07-16 & 1114 & & 15.92 & 15.71 & 23.94 & GIG \\
   F07-17 & 1128 & & 16.20 & 15.91 & 24.40 & (z)GIG \\
   F07-18 & 1139 & & 16.49 & 15.93 & 23.68 & GIG \\
   F07-19 & 1193 & & 16.02 & 15.70 & 23.82 & (z)GIG \\
   F07-20 & 1196 & & 15.82 & 15.46 & 23.65 & (z)GIG \\
   F07-21 & 1281 & & 16.46 & 15.88 & 24.33 & (z)GIG \\
   F07-22 & 1404 & & 15.93 & 15.69 & 23.87 & (z)GIG \\
   F07-23 & 1433 & & 16.19 & 15.74 & 22.23 & GIG \\
   F07-24 & 1532 & & 15.76 & 15.41 & 22.34 & non-GIG/M\\
   F07-25 & 1896 & & 16.42 & 15.79 & 24.09 & GIG\tablenotemark{b} \\
   F07-26 & 1903 & & 15.90 & 15.67 & 22.39 & non-GIG/I\\
   F07-27 & 1908 & & 15.62 & 15.11 & 22.60 & (z)GIG \\
   F07-28 & 1984 & & 15.41 & 15.00 & 21.98 & (z)GIG \\
   F07-29 & 2002 & & 16.45 & 16.17 & 23.43 & (z)GIG \\
   F07-30 & 2074 & & 16.05 & 15.48 & 23.52 & non-GIG/Sp\\
   F07-31 & 2244 & & 15.61 & 15.18 & 24.21 & (z)GIG \\
   \hline
  \end{tabular}
 \end{center}
 \tablecomments{(a) GIG and non-GIG are genuine irregular galaxies and
 not GIG, respectively.  zGIG is the GIG with spectroscopic redshift.  For
 sub-classes of the non-GIG, see Table~\ref{tb:ListOfNonGIGs}.  (b)
 Although F07-25 have spectroscopic information, we do not include
 this object in our zGIG sample since the spectrum appears to be very
 noisy (see footnote 11).}
\end{table*}

\setcounter{table}{1}
\begin{table*}
 \begin{center}
  \caption{A list of 35 irregular galaxies in NA10.}
  \label{tb:ListOfIrrsNA10}
  \begin{tabular}{llccccl}
   \hline\hline
   \multicolumn{2}{c}{Name} & & \multicolumn{2}{c}{Observed Magnitude} &
   \colhead{$\mu_g$} & \colhead{Our Class\tablenotemark{a}}\\
   \cline{1-2} \cline{4-5}
   \colhead{Ours} & \colhead{NA10} & & \colhead{$g$} & \colhead{$r$} &
   $(\mathrm{mag~arcsec^{-2}})$ & \\
   \hline
   NA10-1  & J120301.00$-$001728.21 & & 15.91 & 15.41 & 21.97 & non-GIG/Sp \\
   NA10-2  & J142911.77$-$001415.25 & & 16.05 & 15.51 & 22.67 & non-GIG/I \\
   NA10-3  & J142227.61$+$000332.68 & & 15.98 & 15.16 & 22.04 & non-GIG/I \\
   NA10-4  & J114359.62$+$002540.84 & & 15.92 & 15.51 & 22.27 & non-GIG/Sp \\
   NA10-5  & J135942.73$+$010637.28 & & 16.15 & 15.62 & 21.61 & non-GIG/M \\
   NA10-6  & J014836.90$+$124222.07 & & 16.24 & 15.87 & 21.81 & non-GIG/Sp \\
   NA10-7  & J003234.97$+$150210.45 & & 15.94 & 15.43 & 22.54 & non-GIG/Sp \\
   NA10-8  & J114242.96$-$022138.61 & & 15.93 & 15.60 & 22.32 & non-GIG/Sp \\
   NA10-9  & J085111.80$+$543958.32 & & 15.96 & 15.68 & 22.64 & (z)GIG \\
   NA10-10 & J123012.10$+$033439.88 & & 15.68 & 15.48 & 22.16 & (z)GIG \\
   NA10-11 & J021121.62$-$100715.77 & & 15.89 & 15.57 & 22.81 & (z)GIG \\
   NA10-12 & J020238.79$-$092213.39 & & 16.05 & 15.76 & 22.42 & non-GIG/I \\
   NA10-13 & J225609.41$+$130551.47 & & 15.50 & 15.22 & 21.75 & non-GIG/I \\
   NA10-14 & J211444.78$+$105221.37 & & 16.06 & 15.69 & 20.75 & non-GIG/M \\
   NA10-15 & J223921.86$+$135256.22 & & 15.62 & 15.39 & 22.05 & (z)GIG \\
   NA10-16 & J110037.12$+$040355.07 & & 15.99 & 15.77 & 22.50 & non-GIG/Sp \\
   NA10-17 & J233225.26$-$005049.28 & & 15.98 & 15.67 & 22.54 & (z)GIG \\
   NA10-18 & J024421.10$+$004031.16 & & 16.18 & 15.81 & 23.14 & non-GIG/M \\
   NA10-19 & J114710.16$+$103103.63 & & 15.99 & 15.53 & 22.25 & non-GIG/M \\
   NA10-20 & J162734.31$+$390604.06 & & 16.07 & 15.72 & 21.97 & non-GIG/M \\
   NA10-21 & J014230.89$-$004909.99 & & 15.56 & 15.08 & 22.15 & non-GIG/Sp \\
   NA10-22 & J223436.81$+$001024.37 & & 15.97 & 15.46 & 22.85 & non-GIG/Sp \\
   NA10-23 & J090015.20$+$354319.73 & & 16.48 & 16.46 & 21.71 & (z)GIG \\
   NA10-24 & J092938.47$+$352025.26 & & 15.70 & 15.30 & 23.03 & (z)GIG \\
   NA10-25 & J110508.10$+$444447.09 & & 14.72 & 14.73 & 21.47 & (z)GIG \\
   NA10-26 & J121725.85$+$463400.88 & & 15.02 & 14.79 & 22.37 & non-GIG/M \\
   NA10-27 & J131055.81$+$115229.20 & & 15.82 & 15.57 & 21.87 & non-GIG/M \\
   NA10-28 & J140056.40$+$410025.92 & & 15.64 & 15.33 & 22.77 & non-GIG/I \\
   NA10-29 & J132852.21$+$110549.21 & & 15.92 & 15.56 & 22.29 & (z)GIG \\
   NA10-30 & J113655.36$+$115054.03 & & 15.51 & 15.08 & 21.93 & non-GIG/Sp \\
   NA10-31 & J114830.62$+$124347.08 & & 19.50 & 19.25 & 22.77 & (z)GIG \\
   NA10-32 & J132032.98$+$124922.57 & & 15.76 & 15.42 & 22.02 & non-GIG/M \\
   NA10-33 & J140156.73$+$122249.78 & & 15.78 & 15.57 & 21.92 & (z)GIG \\
   NA10-34 & J131606.19$+$413004.24 & & 15.06 & 14.73 & 21.55 & non-GIG/M \\
   NA10-35 & J102722.42$+$121701.36 & & 16.04 & 15.73 & 22.24 & non-GIG/I \\
   \hline
  \end{tabular}
 \end{center}
 \tablecomments{(a) GIG and non-GIG are genuine irregular galaxies and not
 GIG, respectively.  zGIG is the GIG with spectroscopic redshift.  For sub-classes
 of the non-GIG, see Table~\ref{tb:ListOfNonGIGs}.}
\end{table*}

\setcounter{table}{2}
\begin{table*}
 \begin{center}
  \caption{A list of non-GIGs.}
  \label{tb:ListOfNonGIGs}
  \begin{tabular}{clccccccccccl}
   \hline\hline
   \multicolumn{2}{c}{Contaminant} & \multicolumn{3}{c}{Number} &
   \multicolumn{7}{c}{Details} & \colhead{Sub-class Name} \\
   \cline{3-5}
   & & \colhead{Total} & \colhead{F07} & \colhead{NA10} & & & & & & & & \\
   \hline
   (1) & Elliptical-like galaxies &  0 & 0 & 0     & & & & & & & & non-GIG/E \\
   (2) & Disk-like galaxies       & 11 & 2 & 9     & & & & & & & & non-GIG/Sp \\
       &                          &    &   &       & & Bulge & Spiral
   Arm & Bar & \multicolumn{3}{c}{Number} & \\
   \cline{10-12}
   & & & & & & & & & Total & F07 & NA10 & \\
   \cline{6-12}
   & & & & & (2-a) & {\large $\circ$} & {\large $\circ$} & {\large $\circ$} & 0 & 0 & 0 & \\
   & & & & & (2-b) & {\large $\circ$} & {\large $\circ$} &  $\times$        & 4 & 0 & 4 & \\
   & & & & & (2-c) & {\large $\circ$} &  $\times$        & {\large $\circ$} & 0 & 0 & 0 & \\
   & & & & & (2-d) &  $\times$        & {\large $\circ$} & {\large $\circ$} & 1 & 1 & 0 & \\
   & & & & & (2-e) & {\large $\circ$} &  $\times$        &  $\times$        & 6 & 1 & 5 & \\
   & & & & & (2-f) &  $\times$        & {\large $\circ$} &  $\times$        & 0 & 0 & 0 & \\
   & & & & & (2-g) &  $\times$        &  $\times$        & {\large $\circ$} & 0 & 0 & 0 & \\
   \cline{6-12}
   (3) & Interacting galaxies &  8 & 2 &  6 & & & & & & & & non-GIG/I \\
   (4) & Merging remnants     & 14 & 5 &  9 & & & & & & & & non-GIG/M \\
   \hline
   \multicolumn{2}{c}{Total}  & 33 & 9 & 24 & & & & & & & & \\
   \hline
  \end{tabular}
 \end{center}
\end{table*}

\setcounter{table}{3}
\begin{table*}
 \begin{center}
  \tabletypesize{\scriptsize}
  \tablewidth{\linewidth}
  \caption{A list of 33 GIGs including 27 zGIGs.}
  \label{tb:ListOfGIG}
  \begin{tabular}{llcccccc}
   \hline\hline
   \multicolumn{2}{c}{Name} & \colhead{$z$} & \colhead{$R_\mathrm{P}$}
   &\colhead{$M_g$\tablenotemark{a}} & \colhead{$\mu_g$\tablenotemark{a}} & \colhead{$(g-r)_0$\tablenotemark{a}} &
   \colhead{$12+\log{\mathrm{(O/H)}}$} \\
   \cline{1-2}
   \colhead{Ours} & \colhead{SDSS} &  & (kpc) & (mag) &
		       $(\mathrm{mag~arcsec^{-2}})$ & (mag) & \\
   \hline
   \multicolumn{8}{c}{F07: 22 GIGs including 16 zGIGs} \\
   \hline
   F07-1   & J094407.21$-$003935.3 & ...   & ...  & ...      & 23.00 & ...  & ...  \\
   F07-2   & J094446.23$-$004118.2 & 0.0041& 1.53 & $-15.33$ & 23.42 & 0.30 & 8.13 \\
   F07-3   & J094628.56$-$002603.4 & 0.0060& 1.62 & $-16.53$ & 22.36 & 0.41 & 8.45 \\
   F07-4   & J094705.49$+$005751.2 & 0.0062& 2.02 & $-15.95$ & 23.44 & 0.49 & 8.28 \\
   F07-6   & J102519.78$+$003810.4 & 0.0214& 5.43 & $-19.79$ & 21.74 & 0.28 & 8.62 \\
   F07-7   & J105248.63$+$000203.9 & 0.0060& 1.06 & $-16.72$ & 21.24 & 0.29 & 8.25 \\
   F07-9   & J111054.18$+$010530.5 & 0.0033& 1.79 & $-15.19$ & 23.90 & 0.16 & 8.10 \\
   F07-12  & J112712.26$-$005940.7 & 0.0032& 1.35 & $-15.48$ & 23.01 & 0.39 & 8.32 \\
   F07-13  & J115036.31$-$003403.0 & ...   & ...  & ...      & 23.27 & ...  & ...  \\
   F07-16  & J122021.40$+$002204.2 & ...   & ...  & ...      & 23.94 & ...  & ...  \\
   F07-17  & J122412.46$+$003401.9 & 0.0068& 3.59 & $-16.22$ & 24.40 & 0.26 & 8.06 \\
   F07-18  & J122903.25$+$000616.9 & ...   & ...  & ...      & 23.68 & ...  & ...  \\
   F07-19  & J124002.65$-$010257.6 & 0.057 & 2.36 & $-15.88$ & 23.82 & 0.29 & 8.22 \\
   F07-20  & J124008.77$-$002107.7 & 0.066 & 4.84 & $-16.20$ & 23.65 & 0.34 & 8.35 \\
   F07-21  & J125405.16$-$000604.3 & 0.0032& 1.46 & $-14.32$ & 24.33 & 0.55 & 8.81 \\
   F07-22  & J131743.19$-$010003.8 & 0.0041& 1.95 & $-15.42$ & 23.87 & 0.21 & 8.30 \\
   F07-23  & J132009.17$-$011128.3 & ...   & ...  & ...      & 22.23 & ...  & ...  \\
   F07-25  & J144300.18$-$002300.2 & (0.8124)\tablenotemark{b} & (156.6) & $(-27.28)$ & 24.09 & ... & ... \\
   F07-27  & J144515.80$-$000934.3 & 0.0056& 1.72 & $-16.41$ & 22.60 & 0.47 & 8.41 \\
   F07-28  & J150001.30$-$010527.8 & 0.0063& 1.68 & $-17.00$ & 21.98 & 0.34 & 8.35 \\
   F07-29  & J150350.19$+$005841.8 & 0.0070& 2.20 & $-16.14$ & 23.43 & 0.23 & 8.20 \\
   F07-31  & J154219.30$+$002831.3 & 0.0066& 4.84 & $-17.06$ & 24.21 & 0.32 & 8.22 \\
   \hline
   \multicolumn{8}{c}{NA10: 11 GIGs including 11 zGIGs}\\
   \hline
   NA10-9  & J085111.76$+$543958.3 & 0.027 & 6.01 & $-19.39$ & 22.64 & 0.33 & 8.36 \\
   NA10-10 & J123012.10$+$033439.9 & 0.018 & 3.82 & $-18.80$ & 22.16 & 0.20 & 8.29 \\
   NA10-11 & J021121.60$-$100716.3 & 0.013 & 5.30 & $-17.99$ & 22.81 & 0.32 & 8.31 \\
   NA10-15 & J223922.29$+$135300.2 & 0.018 & 3.30 & $-19.48$ & 22.05 & 0.25 & 8.30 \\
   NA10-17 & J233225.25$-$005049.1 & 0.018 & 4.11 & $-18.58$ & 22.54 & 0.32 & 8.41 \\
   NA10-23 & J090015.20$+$354319.7 & 0.011 & 3.66 & $-18.05$ & 21.71 & 0.21 & 8.08 \\
   NA10-24 & J092938.46$+$352025.2 & 0.015 & 5.89 & $-18.26$ & 23.03 & 0.36 & 8.51 \\
   NA10-25 & J110508.11$+$444447.1 & 0.022 & 3.62 & $-19.99$ & 21.47 & 0.02 & 8.27 \\
   NA10-29 & J132852.21$+$110549.1 & 0.015 & 3.28 & $-18.12$ & 22.29 & 0.34 & 8.38 \\
   NA10-31 & J114830.64$+$124347.6 & 0.013 & 4.66 & $-17.76$ & 22.77 & 0.26 & 8.50 \\
   NA10-33 & J140156.74$+$122249.7 & 0.023 & 3.79 & $-19.36$ & 21.92 & 0.20 & 8.35 \\
   \hline
%   \tablenotetext{1}{Since the spectrum appears to be very noisy, we do not include this object in our zGIG sample.}
%   \tablenotetext{2}{Rest-frame $g-r$ color.}
 \end{tabular}
 \end{center}
 \tablecomments{(a) The Galactic extinction correction is performed.
 (b) Since the spectrum appears to be very noisy, we do not include
 this object in our zGIG sample.}
\end{table*}

%%%%%%%%%%%%%%%%%%%%%%%%%%%%%%%%%%%%%%%%%%%%%%%%%%%%%%%%%%%%%%%%%%%%%%%%%%%%%
\appendix

 \section{A.  Comments on Each Galaxy Re-Classified into the
 Non-GIGs/I$+$M}\label{sec:appA}

 Here we provide comments on the irregular galaxies in the preliminary
 samples identified as either interacting or merging galaxies in our
 classification.

 \begin{enumerate}
  \item Interacting galaxies
	\begin{description}
	 \item[F07-8] An asymmetric structure is seen in NE.
	 \item[F07-26] This object is regarded as a ring galaxy (Theys
		    \& Spiegel 1976; Toomre \& Toomre 1972).  A
		    probable colliding partner is a small galaxy seen
		    at SSE.
	 \item[NA10-2] Weak but asymmetric feature can be seen. A
		    small galaxy seen at NEE may be an interacting
		    partner.
	 \item[NA10-3] An iterating partner can be seen at W.
	 \item[NA10-12] The overall structure is highly asymmetric. An
		    interacting partner is located at SSE.
	 \item[NA10-13] The overall structure is asymmetric. An
		    interacting partner is located at N.
	 \item[NA10-28] The overall structure is asymmetric. An
		    interacting partner is located at S.
	 \item[NA10-35] The overall structure is asymmetric. An
		    interacting partner is located at SW.
	\end{description}
  \item Merging galaxies
	\begin{description}
	 \item[F07-5] One-sided arm can be seen at N. This structure
		    may take an polar orbit, being similar to Arp 336
		    (NGC 2685; Arp 1966). A small object located at S
		    may be a merging partner or its relic.
	 \item[F07-10] The overall structure is highly asymmetric,
		    evidenced by extended structure at E.
	 \item[F07-14] One-sided structure is emanated from the
		    southern part of the main body.  This can be
		    regarded as a tidal tail.
	 \item[F07-15] One-sided structure is emanated from the NW
		    part of the main body.  This can be regarded as a
		    tidal tail.
	 \item[F07-24] One-sided structure is emanated from the
		    northern part of the main body. This can be
		    regarded as a tidal tail.
	 \item[NA10-5] One-sided structure is emanated from the
		    eastern part of the main body.  This can be
		    regarded as a tidal tail.
	 \item[NA10-14] One-sided structure is emanated from the NE
		    part of the main body.  This can be regarded as a
		    tidal tail.
	 \item[NA10-18] The overall structure is asymmetric, evidenced
		    by extended structure toward N.
	 \item[NA10-19] One-sided structure is emanated from the SE
		    part of the main body.  This can be regarded as a
		    tidal tail seen from the edge-on.
	 \item[NA10-20] The overall structure is highly asymmetric,
		    evidenced by small (at E) and large (W) plume-like
		    structures.
	 \item[NA10-26] The overall structure shows significant bending.
	 \item[NA10-27] The overall structure shows significant bending.
	 \item[NA10-32] The overall structure shows an integral-like
		    pattern.
	 \item[NA10-34] The overall structure shows an inverse
		    integral-like pattern.  The small knot located at
		    the NNW edge may be a merging partner.
	\end{description}
 \end{enumerate}

 \section{B.  Comments on the GIGs with Apparent
 Companion}\label{sec:appB}

 Here we give comments on the following four GIGs with apparent
 companion in their images shown in Figure~\ref{fig:GIGs}: F07-22,
 F07-23, NA10-23, and NA10-29.

 \begin{description}
  \item[F07-22] The object at the NE edge of the image in
	     Figure~\ref{fig:GIGs} is classified as a star in the SDSS
	     DR10.
  \item[F07-23] The yellow extended object at NW of the F07-23 is a
	     galaxy at $z = 0.11$.  Although F07-23 does not have
	     spectroscopic information, it may be located at a similar
	     redshift with the F07-zGIGs ($= 0.0032$--0.027), which is
	     much smaller than that of the NW galaxy.  Note that no
	     signature of interaction is seen in the NW galaxy albeit
	     its relatively high surface brightness.
  \item[NA10-23] There is a large blue extended object at N of
	     NA10-23.  It is classified as a galaxy in the SDSS DR10
	     and does not have a spectroscopic information.  Its
	     photometric redshift is highly uncertain and includes the
	     redshift of NA10-23 ($= 0.011$) within 1$\sigma$ error.
	     However, no signature of interaction is seen in the N
	     galaxy albeit its relatively high surface brightness ($=
	     21.75~\mathrm{mag~arcsec^{-2}}$).
  \item[NA10-29] The yellow extended object at NNW of the NA10-29 is
	     classified as a galaxy in the SDSS DR10.  Although it
	     does not have spectroscopic information, its photometric
	     redshift is provided as $\approx 0.06$, which is much
	     larger than the redshift of NA10-29 ($= 0.015$).
 \end{description}
%%%%%%%%%%%%%%%%%%%%%%%%%%%%%%%%%%%%%%%%%%%%%%%%%%%%%%%%%%%%%%%%%%%%%%%%%%%%%

%% %%%%%%%%%%%%%%%%%%%%%%%%%%%%%%%%%%%%%%%%%%%%%%%%%%%%%%%%%%%%%%%%%%%%%%%%%%%

\end{document}